\documentclass[final,5p,times,twocolumn,authoryear]{elsarticle}
\usepackage{amssymb}
\usepackage{lipsum}

\journal{Astronomy $\&$ Computing}

\usepackage{hyperref}
\usepackage{xcolor}
\usepackage{graphicx}	
\usepackage{amsmath}	
\usepackage[final]{listings}
\usepackage{xspace}
\usepackage{multicol}
\usepackage[frozencache,cachedir=.]{minted}

\usepackage[T1]{fontenc}

\newcommand{\sect}[1]{\S\ref{#1}}

\newcommand{\dask}{\mbox{\textsc{Dask}}}

\newcommand{\daskms}{\mbox{\textsc{Dask-MS}}}

\newcommand{\zarr}{\mbox{\textsc{Zarr}}}
\newcommand{\apache}{\mbox{\textsc{Apache-Arrow}}}
\newcommand{\cubical}{\mbox{\textsc{CubiCal}}}
\newcommand{\quartical}{\mbox{\textsc{QuartiCal}}}
\newcommand{\pfbclean}{\mbox{\textsc{pfb-imaging}}}
\newcommand{\ddfacet}{\mbox{\textsc{DDFacet}}}
\newcommand{\wsclean}{\mbox{\textsc{WSClean}}}

\newcommand{\oldstimela}{\mbox{\textsc{Stimela}}}
\newcommand{\stimela}{\mbox{\textsc{Stimela2}}}
\newcommand{\omegaconf}{\mbox{\textsc{OmegaConf}}}
\newcommand{\cultcargo}{\mbox{\textsc{cult-cargo}}}
\newcommand{\cultCargoUrl}{\url{https://github.com/caracal-pipeline/cult-cargo}}
\newcommand{\caracal}{{\sc CARACal}}
\newcommand{\vermeerkat}{{\sc VermeerKAT}}
\newcommand{\slurm}{{\sc slurm}}
\newcommand{\kubernetes}{{\sc Kubernetes}}
\newcommand{\knetes}{{\sc K8s}}
\newcommand{\simms}{{\sc simms}}
\newcommand{\telsim}{{\sc telsim}}
\newcommand{\skysim}{{\sc skysim}}
\newcommand{\oxkat}{{\sc OxKAT}}
\newcommand{\daliuge}{{\sc DALiuGE}}

\newcommand{\red}[1]{\textcolor{red}{#1}}

\DeclareRobustCommand{\VAN}[3]{#2}
\let\VANthebibliography\thebibliography
\def\thebibliography{\DeclareRobustCommand{\VAN}[3]{##3}\VANthebibliography}

\definecolor{yamlGreen}{RGB}{63,127,63}
\definecolor{yamlBlue}{RGB}{0,0,255}
\definecolor{yamlPurple}{RGB}{127,0,127}
\definecolor{yamlGray}{RGB}{100,100,100}

\lstdefinestyle{yaml}{
    language={},
    basicstyle=\ttfamily\scriptsize,
    numbers=left,
    numberstyle=\tiny\color{gray},
    keywordstyle=\color{yamlGreen},
    stepnumber=1,
    numbersep=5pt,
    backgroundcolor=\color{gray!10},
    showspaces=false,
    showstringspaces=false,
    showtabs=false,
    frame=single,
    tabsize=2,
    captionpos=b,
    breaklines=true,
    breakatwhitespace=true,
    breakautoindent=true,
    linewidth=\textwidth,
    moredelim=[l][\color{yamlGray}]{\#},
    moredelim=[l][\color{yamlPurple}]{---},
    moredelim=**[s][\color{yamlBlue}]{*}{*},
    morestring=[b]',
    morestring=[b]"
}

\begin{document}
\begin{frontmatter}



\title{Africanus IV. The Stimela2 Framework: Scalable And Reproducible Workflows, From Local To Cloud Compute}

\begin{abstract}
    \stimela\ is a new-generation framework for developing data reduction workflows. It is designed for radio astronomy data but can be adapted for other data processing applications. \stimela\ aims at the middle ground between ease of development, human readability, and enabling robust, scalable and reproducible workflows. It represents workflows by linear, concise and intuitive YAML-format \emph{recipes}. Atomic data reduction tasks (binary executables, Python functions and code, and CASA tasks) are described by YAML-format \emph{cab definitions} detailing each task's \emph{schema} (inputs and outputs). \stimela\ provides a rich syntax for chaining tasks together, and encourages a high degree of modularity: recipes may be nested into other recipes, and configuration is cleanly separated from recipe logic. Tasks can be executed natively or in isolated environments using containerization technologies such as Apptainer. The container images are open-source and maintained through a companion package called \cultcargo. This enables the development of system-agnostic and fully reproducible workflows. \stimela\ facilitates the deployment of scalable, distributed workflows by interfacing with the {\sc Slurm} scheduler and the \kubernetes\ API. The latter allows workflows to be readily deployed in the cloud. Previous papers in this series used \stimela\ as the underlying technology to run workflows on the AWS cloud.

    This paper presents an overview of \stimela's design, architecture and use in the radio astronomy context.

\end{abstract}
    
\author[ratt,sarao,ira]{O.~M.~Smirnov}\ead{o.smirnov@ru.ac.za}
\author[wits,ratt,oac]{S.~Makhathini}
\author[ratt]{J.~S.~Kenyon}
\author[sarao,ratt]{H.~L.~Bester}
\author[sarao]{S.~J.~Perkins}
\author[sarao,ratt]{A.~J.~T.~Ramaila}
\author[sarao,ratt]{B.~V.~Hugo}

\affiliation[ratt]{organization={Centre for Radio Astronomy Techniques \& Technologies (RATT),
Department of Physics and Electronics, Rhodes University},
            city={Makhanda},
            state={EC},
            country={South Africa}}

\affiliation[sarao]{organization={South African Radio Astronomy Observatory (SARAO)},
            city={Cape Town},
            state={WC},
            country={South Africa}}

\affiliation[ira]{organization={Institute for Radioastronomy, National Institute of Astrophysics (INAF IRA)},
            city={Bologna},
            country={Italy}}

\affiliation[wits]{organization={School of Physics, University of the Witwatersrand},
            city={Johannesburg},
            state={Gauteng},
            country={South Africa}}

\affiliation[oac]{organization={Observatory of Cagliari, National Institute of Astrophysics (INAF OAC)},
            city={Cagliari},
            country={Italy}}


\end{frontmatter}



\section{Introduction}

Radio astronomy data reduction has been anecdotally described as ``death by a million papercuts''. The past decade has exacerbated this, with several new radio facilities coming online, each with its own instrumental quirks and data challenges, and with rapid progress in algorithmic and software developments. Radio astronomy packages are becoming increasingly arcane -- for example, packages such as \wsclean\ \citep{wsclean} and \ddfacet\ \citep{ddfacet} have many dozens of parameters -- with relatively few ``black belt'' experts possessing the expertise 
to exploit them optimally. On the other hand, building ``black box'' data reduction 
pipelines aimed at the relatively unqualified user has, historically, proven challenging \citep[see discussion of this problem by][]{molenaar-phd}. There are some success stories: the ALMA and VLA pipelines provided by CASA serve the needs of most users for standard observations. The word ``standard'', however, is an important qualifier: the sensitivity, high spatial resolution and field-of-view of new instruments leads to new and more subtle calibration and imaging problems that require non-standard, often novel, software tools. Now, while such tools (e.g, \wsclean, \ddfacet, \pfbclean, \quartical) are available and continue being developed, incoperating them into pipelines requires expert knowledge of both the tools and the pipelines, making it prohibitively difficult for most astronomers. The \oxkat\ pipeline \citep{oxkat} integrates a number of these novel tools into a single workflow, and ships as a single Docker image with all software pre-installed. The \caracal\ pipeline \citep{caracal} attempts to address this by using the first-generation \oldstimela\ package \citep{stimela} as the backend, and providing a YAML-format user configuration. \caracal\ is somewhat more transparent and includes a rich set of legacy and novel tools, but the process of adding new tools is verbose and cumbersome due to limitations of the \oldstimela\ backend (discussed below). While both of these pipelines have done a great deal to empower data reduction by non-expert MeerKAT users, they do not readily lend themselves to casual user modification. 
Under such conditions, it remains the case that most high-fidelity data reductions are the result of handcrafted data reduction by black belts (a process also referred to in the community as ``hero mode'' data reduction). This also contributes to a reproducibility crisis -- not unique to radio astronomy -- in that while the raw observational data for any particular science result may be publicly available, the data reduction leading to it may not be readily reproducible. Minor changes in parameter settings and package versions can often lead to appreciable changes in the output images.

This creates an important problem: how do we simultaneously manage software complexity, make diverse (and rapidly evolving) software packages work together, create fully reproducible data reduction recipes\footnote{\emph{Recipes}, in the fully general sense of procedures and best practices. The term \emph{recipe} will take on a more specific technical meaning later in this work.}, while allowing black belt users, who in the normal course of their job develop new recipes, to share these with the community in an accessible way? Finally, an issue that ends up touching upon all of these questions, is, as we'll see later in this work, how (and if) can we use cloud computing resources to run such recipes?

\paragraph*{Cloud computing in astronomy} Cloud computing has become a mainstream technology, and it is a common choice for businesses to deploy their data processing on commercial clouds, such as those provided by Amazon (AWS), Google (GCE) and Microsoft (Azure). Simple economy-of-scale arguments suggest that big commercial providers must be able to supply computing resources at lower unit cost than in-house compute, for all but the biggest organizations -- the caveat is, does a given compute workflow and software stack map onto the cloud compute model efficiently? In astronomy, the flagship example is provided by the Rubin Observatory, which is deploying its entire Science Data Platform\footnote{\url{https://data.lsst.cloud}} on the Google cloud
\citep{rubin2023}; see also \citet{berriman-cloud} for a discussion. At 100+ Petabytes in scale, the Rubin SDP project is in the same class as the SKA and its precursors. Despite a concerted effort by AWS and the SKA Observatory (dating back to 2015) to promote cloud development in radio astronomy\footnote{\url{https://aws.amazon.com/blogs/aws/new-astrocompute-in-the-cloud-grants-program/}}, our community has been relatively slow to exploit this technology, although a few project-specific demonstrations and deployments have been reported \citep{cloudpulsar2017,cloudlofar2017,cloud21cmcosmology2021,dodson-cloud,dodson-cloud2}. 

This reluctance could be due to both sociological reasons and technological bottlenecks. Of the latter, we can identify at least two that are relevant to this paper series. Firstly, our most established data formats, in particular the Measurement Set and its underlying Casacore Table Data System \citep{Diepen2015} require a POSIX filesystem, which, in a cloud environment, necessitates the use of the more expensive \emph{block store} (as opposed to cheaper \emph{object store} options such as S3). The format is also ill-suited for parallel I/O, as discussed earlier in this paper series by \citet{africanus1}. Storage costs are also exacerbated by legacy tools that commonly at least double or triple the on-disk data volumes (viz., CASA's practice of creating model and corrected data columns.)
Secondly, our data reduction workflows, besides being long and complex, tend to be ``thick-thin'', in the sense that some steps may be highly parallelizable (e.g. gridding), while others are strictly serial (e.g. classic CLEAN minor cycle); some have a large RAM footprint, while others are very economical. This makes it particularly challenging to utilize cloud resources in a cost-efficient way, since one is charged an hourly rate for a given virtual machine instance, regardless of actual CPU or RAM utilization. This paper series discusses our solution for both of these bottlenecks.

\paragraph*{Containerization} \label{oldstimela} The original \oldstimela\ package aimed to address reproducibility and ease-of-installation concerns by implementing a \emph{containerized} workflow framework. \oldstimela\ provides a set of curated container images for most of the popular radio astronomy packages (these images being regularly updated with each release of \oldstimela). Each image is complemented by a JSON-format \emph{schema} formally describing the inputs and outputs of the corresponding package. Together, the image and the schema is known as a \emph{cab} definition. These cabs can be chained together via a Python API to produce \emph{recipes}, the latter being a sequence of steps, with each step invoking a particular cab with a set of parameters. \oldstimela\ takes care of parameter validation (attempting to catch as many errors up front as possible), and then executes the recipe, by instantiating the images as Docker (or Apptainer/Singularity, or Podman) containers and passing them the appropriate parameters. \oldstimela\ addressed some of the above concerns: wrapping the packages into container images takes away most of the software installation complexity, and allows the recipe to run virtually anywhere (only Python, and Docker or Podman or Apptainer is needed on the host system) in a reproducible manner. \oldstimela\ has had success as the underlying engine for the 
\vermeerkat\footnote{\url{https://github.com/ratt-ru/vermeerkat}} and \caracal\ pipelines. We have also observed a lot of casual use in the community, although primarily to run individual cabs in one-off mode (thus using \oldstimela\ as, essentially, a software distribution tool), but there has been no meaningful recipe development outside the core team. 
We have identified three major shortcomings in the original design: (i) recipe execution is strictly serial, which prevents cabs without interdependence from running in parallel. Any more complex flow control (or parallelization) is deferred to the calling Python code, however (ii) the Python-based API has proven ill-suited as a standardized recipe-building framework. In particular, long and complicated workflows tend to devolve to, essentially, regular Python scripts, and become hard to read -- which goes against the core aim of simplifying recipe-building. Finally, (iii) while invoking native binary applications from within a recipe can be done via standard Python modules, this effectively bypasses \oldstimela\ entirely. This limits rapid development and experimentation in a research environment.

\stimela\ attempts to address these shortcomings through a number of new features:

\begin{itemize}
    \item A concise, rich, and human-friendly YAML-based recipe syntax, replacing the Python API;
    \item High degree of portability: libraries of recipes and custom cab definitions can be distributed as YAML files;
    \item A rich formula and string substitution syntax for specifying parameter values;
    \item Support for scatter-gather constructs within a recipe;
    \item Cabs are no longer restricted to container images and Python functions: native binaries and virtual environments can be wrapped into cabs, and intermixed within a single workflow;
    \item Support for distributing workflows over \slurm\footnote{\url{https://slurm.schedmd.com/}} and \kubernetes\footnote{\url{https://kubernetes.io/}} clusters;
    \item Built-in performance metrics and profiling.
\end{itemize}

\paragraph*{Dataflow programming versus scripting} The first paper of this series \citep{africanus1} already introduced dataflow programming (DFP) in the context of high-performance numeric code -- i.e., relatively low-level -- implementations. DFP has also gained traction at the higher level, that of workflow management. The advent of Big Data in many areas of science has driven demand for developing and deploying scalable and reproducible workflows.
An example of applying DFP to this problem
is the Common Workflow Language\footnote{\url{https://www.commonwl.org}} (CWL), which originated in the biomedical and bioinformatics field, but has since been adopted within the LOFAR project\footnote{\url{https://github.com/EOSC-LOFAR/prefactor-cwl}}. \daliuge\ \citep{daliuge1, daliuge2} is a DFP framework that was specifically developed to address the large-scale data processing requirements of SKA1.  

In this context, DFP and scripting are, in a sense, at opposite ends of a spectrum. DFP is very well-suited to defining highly scalable, parallelized workflows, as demonstrated by \citet{daliuge-scaling}. It does, however, require somewhat specialized domain expertise: 
we are not aware of any casual astronomer-users (even black belt ones) readily developing their own custom workflows with CWL or \daliuge.
Scripting is at the other extreme: it requires only minimal programming skills, can be casually adopted by any astronomer, but the resulting code hardly promotes scalability, portability or reproducibility. 

\paragraph*{Aims} \stimela\ aims to exist somewhere in the middle of this spectrum. \stimela\ workflows are described by script-like linear recipes, easily accessible to the casual astronomer-programmer, yet with advanced constructs such as scatter-gather enabling scalability. 
The \stimela\ \emph{cab definition} scheme allows existing packages to be wrapped into ``black boxes'' (and ultimately containerized) non-invasively and with minimal effort, without involving the package developers. At the same time, recipe and cab definitions are sufficiently formalized that \stimela\ is able to turn them into a dataflow-like construct under the hood, and scale them up on e.g. a \kubernetes\ or \slurm\ cluster. The goal of \stimela\ is to provide a workflow framework that is (a) easy to develop in, (b) deploys locally in a development environment with minimum fuss, (c) can provide full reproducibility and is (d) reasonably performant and scalable when deployed in an HPC environment. As we shall see later in this work, meeting these goals enables (e) the deployment of \stimela\ workflows in a cloud environment.

Previous papers in this series \citep{africanus2,africanus3} demonstrated how to use \stimela\ to deploy and drive scalable workflows, including in a cloud environment. In the meantime, the RATT PARROT paper \citep{parrot} has been published along with a complete set of \stimela\ recipes for the (somewhat non-standard) data reduction procedures used in that work. These recipes seamlessly combine multiple software packages and snippets of Python code, and should be fully reproducible by anybody, given a sufficiently large compute node with only \stimela\ and Apptainer/Singularity installations, or, alternatively, access to an AWS EKS cluster.

The aim of this paper is to present the \stimela\ design and its application to radio astronomy workflows. This is not a user guide, but rather an introduction and whistlestop tour of the major features of the framework.
We therefore strive to avoid excessive technical detail --  for the latter, the reader is urged to refer to the official documentation page.\footnote{\url{https://stimela.readthedocs.io}}

\section{Main concepts}
\label{sec:mainconcepts}

\paragraph*{Recipes} At the most basic level, a workflow in \stimela\ is represented by a \emph{recipe}. A recipe has a set of inputs and outputs (defined by its \emph{schema}), and contains a sequence of steps. Each step either invokes a \emph{cab}, or another recipe (which is then referred to as a sub-recipe), specifying a number of \emph{parameter} values that are validated against the cab's or sub-recipe's schema. The command 

{\small\begin{verbatim}
  $ stimela run recipe.yml myrecipe ms=foo.ms
\end{verbatim}}

\noindent tells \stimela\ to load recipe definitions from the file {\tt recipe.yml}, and run a recipe from therein called {\tt myrecipe}, setting the input named {\tt ms} to the value {\tt "foo.ms"} (Listing~\ref{fig:myrecipe}, but see also \ref{app:telsim} for a real-life example).

\begin{listing}
\begin{minted}[fontsize=\small]{yaml}
myrecipe:
  inputs:
    foo:
      dtype: str
      default: "this is the default setting for foo"
    ms: 
      dtype: MS
      required: true
  steps:
    img:
      cab: wsclean
      params: 
        ms: =recipe.ms
        size: [1024, 1024]
    cal:
      cab: cubical
      params:
        ms: =recipe.ms
\end{minted}
\caption{A very simple and notional {\tt recipe.yml}.}
    \label{fig:myrecipe}
\end{listing}

This notional example demonstrates some fundamental concepts: a recipe's schema defines zero or more inputs and outputs, a recipe contains a sequence of arbitrarily-named steps (here, {\tt cal} and {\tt img}), each step invokes a cab ({\tt cubical} and {\tt wsclean}), and at each step we specify the cab's inputs and outputs via a {\tt params} section. Finally, parameters can reference other parameters (here, {\tt =recipe.ms} means ``use the value of the {\tt ms} parameter of the parent recipe'') in various interesting ways (\sect{sec:subst}).

The example recipe also illustrates the fundamental data structure employed throughout the \stimela\ architecture: the ordered dictionary (or {\em mapping}, in YAML parlance). An ordered dictionary is a set of keys and values; keys are unique; key order is meaningful ({\tt img} comes before {\tt cal}). Most constructs in \stimela\ are represented by such nested ordered dictionaries, which naturally map to the YAML syntax of indented sections. A recipe is a nested ordered dictionary containing keys (a.k.a. sections) such as {\tt inputs}, {\tt outputs} and {\tt steps}; the latter is an ordered dictionary where each step's name is the key, and the value is an ordered dictionary defining the step, and so on all the way down. We will use the term {\em section} to refer to keys whose values are nested dictionaries, and {\em field} for keys whose values are primitive types. This terminology intuitively matches the visual layout of the YAML files. 

\paragraph*{Cabs} Cabs are the atomic tasks available to \stimela. A cab has a defined set of inputs and outputs given by its \emph{schema}. The \emph{cargo} (i.e. content) of a cab can be a any executable shell command, a Python function, a CASA task, or indeed any snippet of Python code. Depending on the chosen \emph{backend}, cabs can be executed natively on the host system, remotely, and/or as containers.
Listing~\ref{fig:cab} is an example cab definition, corresponding to the {\tt shadems} command.
The {\tt inputs} and {\tt outputs} sections define a cab's parameter schema. 

Unlike its first-generation precursor, \stimela\ itself does not include any standard radio astronomy cabs. 
Instead, it is designed to work with user-installable (and/or user-defined) collections of cabs. There is 
an ``official'' collection of cab definitions called \cultcargo\footnote{\cultCargoUrl} which is distributed via PyPI. This provides cabs for most popular radio astronomy packages, as well as for individual CASA tasks. It can be loosely thought of as the KERN suite\footnote{\url{https://kernsuite.info}} for \stimela\ (at least once it is reasonably complete -- the current public release is a subset, but the suite is under active development). A corresponding set of versioned container images is uploaded to the {\tt https://quay.io} registry, from where \stimela\ can automatically download them on demand (that is, the first time a cab is invoked) if a containerized backend is employed. Note that \cultcargo\ is only ``official'' in the sense that it is maintained directly by the \stimela\ team; other users or teams are free to create and distribute their own cab collections, and these can be intermixed in any given workflow (see \ref{app:telsim} for an example).

In addition to this, the user can define arbitrary custom cabs, by writing YAML similar to Listing~\ref{fig:cab}. The use-and-include features (\sect{sec:use-include}) allow these definitions to be structured into modular, reusable and shareable libraries of cab definitions.

\paragraph*{Schemas} A schema defines the inputs and outputs of a cab (or recipe, as we saw above). These play a crucial role in \stimela. A schema is defined as a list of named sections, one per parameter, containing a data type specifier (the {\tt dtype} field), plus help strings, policies, and other optional attributes. The type specifier uses the same syntax as the standard Python {\tt typing} module, which allows for a rich variety of data types, including compound constructs such as

{\small
\begin{verbatim} 
  Union[int, str, Tuple[int, int]]
\end{verbatim}
}

\noindent ...meaning, in this case, that a parameter may be an integer, a string, or two integers. \stimela\ extends this with a few additional types, such as {\tt File}, {\tt Directory} and {\tt MS}. It is also possible to specify a {\tt choices} field, for a parameter than can only assume a restricted set of values.

\paragraph*{Parameter validation} The type information in the schema allows \stimela\ to perform fairly extensive parameter validation. Our encouraged approach is to specify fairly strict schemas in the cabs, so as to catch as many user errors as possible (though developers remain free to defeat this purpose by employing permissive types such as {\tt str} or {\tt Any}). Validation is done in two stages. {\em Prevalidation} is performed before the recipe is executed; this checks the recipe and all steps for any missing required parameters, incorrect types for known parameters, missing files (for {\tt File} and {\tt MS}-type parameters), etc. This attempts to catch as many errors as possible up front. Additional \emph{runtime validation} is done before and after each step. This catches errors in parameter values that arise at runtime (particularly in cases where the output of one step is used as the input of another step).

\paragraph*{Substitutions and evaluations} These are performed on parameter values at runtime. We saw an example in the notional recipe, with {\tt"=recipe.ms"} being used to refer to the {\tt ms} input of the parent recipe. Any parameter value that starts with an ``='' sign is treated as a formula \emph{evaluation}. For example, {\tt "=current.x + 1"} is evaluated as the value of the parameter {\tt x} of the current step, plus one. The formula language is explained in more detail in \sect{sec:subst}. {\em Substitutions} on strings are invoked using the curly braces construct (inspired by and implemented via Python format strings). Using something like {\tt "new-\{recipe.image\_name\}.fits"} would result in taking the value of the {\tt image\_name} parameter of the recipe, and adding the prefix {\tt "new-"} and the suffix {\tt ".fits"}. This is a very powerful mechanism for keeping the ``million papercuts'' problem under control. A typical workflow will generate dozens of data products, and will have fairly complex parameter interactions between steps. Using substitutions and evaluations makes it much easier to keep track of parameters, and to enforce naming conventions for output files.

\begin{listing}
\begin{minted}[fontsize=\small]{yaml}
name: shadems
image: 
  name: shadems
command: shadems
info: 'Rapid Measurement Set plotting with xarray-ms 
  and datashader'

defaults:
  norm: auto
  xcanvas: 1280
  ycanvas: 900

policies:
  prefix: '--'
  positional: false
  repeat: list

inputs:
  ms:
    info: Measurement set to plot
    required: true
    dtype: MS
    policies:
      positional: true
  xaxis:
    info: 'X axis to plot. Can be any MS column name, 
      also: CHAN, FREQ, CORR, ROW, WAVEL, U, V, W, UV, 
      and, for complex columns, keywords such as amp, 
      phase, real, imag. You can also specify 
      correlations, e.g. DATA:phase:XX. The order of 
      specifiers does not matter.'
    dtype: str
  yaxis:
    info: 'Y axis to plot. Must be given the same 
      number of times as xaxis.'
    dtype: str
  ...
    \end{minted}
    \caption{Example of a cab definition.}
    \label{fig:cab}
\end{listing}

\subsection{Running recipes and accessing documentation}

As seen above, the basic way to run a recipe is to specify a YAML file, a recipe name, and the required parameters, if any:

{\small
\begin{verbatim} 
  $ stimela run recipe.yml myrecipe ms=foo.ms
\end{verbatim}
}

If the document contains a single recipe only, the recipe name is not necessary. Alternatively, {\tt run -l} runs the \emph{last} recipe in a document.

Running 

{\small
\begin{verbatim} 
  $ stimela doc recipe.yml 
\end{verbatim}
}

\noindent prints documentation on the recipe and its parameters (using the embedded {\tt info} strings which, hopefully, have been provided by the recipe developer).

The {\tt doc} command can also document cabs. For example, assuming the {\tt cult-cargo} package is installed, 

{\small
\begin{verbatim} 
  $ stimela doc cultcargo::wsclean.yml
\end{verbatim}
}

\noindent will print complete documentation on the \wsclean\ cab (note that the ``{\tt ::}'' syntax on the command line is shorthand for ``find Python module named {\tt cultcargo} and look for the specified YAML file therein'').

Finally, note that both {\tt run} and {\tt doc} will accept multiple YAML documents and \emph{compose} (see next section) them together before looking at the recipe. This makes it straightforward to separate the logical recipe per se from, say, local configuration tweaks, and compose the two with:

{\small
\begin{verbatim} 
  $ stimela run recipe.yml config.yml myrecipe
\end{verbatim}
}

\section{YAML, nested namespaces and composability}
\label{sec:global-namespace}
The nested dictionary, which we also call the nested namespace (and YAML calls a mapping), is the primary data structure of \stimela. Pretty much everything in the system is represented by a nested dictionary, including the overall system configuration (a.k.a. the {\em config namespace}). At the top level, the config namespace has the following structure:

\begin{minted}[fontsize=\small]{yaml}
cabs:        # cab definitions
  foo:       #   defines cab foo
    # ...
  bar:       #   defines cab bar
    # ...
    
lib:         # libraries of reusable components
  recipes:   #   library of recipes
    my_recipe: 
      # ...
  # other free-form library entries

vars:        # arbitrary user-defined variables
  x: 0
  y: "a string"

opts:        # runtime options
  log:       #   logging options
    # ...
    
run:         # runtime environment information
  date:      #   date the run started
  time:      #   time the run started
  node:      #   hostname of current node
  env:       #   environment variables from shell
    PATH: 
    # ...
\end{minted}

The config namespace is populated at startup, firstly from (a) the \stimela\ configuration, (b) runtime information (in particular, for the {\tt run} namespace), and finally (c) user-specified configuration and recipe files, which define recipes and include (see below) cab definitons.

\subsection{YAML extensions: use, include, merge and augment}
\label{sec:use-include}

\newcommand{\ttinclude}{{\tt \_include}}
\newcommand{\ttuse}{{\tt \_use}}

Bringing configuration information together from a large set of sources (YAML files) has necessitated some additional design features to enable modularily and re-use of YAML. The YAML standard\footnote{\url{https://yaml.org/spec/1.2.2}} offers some rather rudimentary features (anchors, aliases and merge) in support of this. A package called \omegaconf\footnote{\url{https://github.com/omry/omegaconf}} provides a number of very useful extensions such as variable interpolation, as well as a very convenient API for working with
large YAML configurations and mapping them to Python dataclasses. This package was adopted by \stimela\ as the underlying technology. 
On top of this, \stimela\ defines a module called {\tt configuratt} which implements a number of useful YAML extensions supporting modularity and libraries. In particular, these are the \ttinclude\ and \ttuse\ keys. 

The core idea behind these constructs is very simple, and familiar to any programmer: they allow for snippets of YAML to be defined in one place, and reused elsewhere, library-style. An \ttinclude\ key \emph{merges} the entire content of one or more YAML file(s) into the section where it appears. The \ttuse\ key merges the content of one or more previously defined sections into the  section where it appears. All other content in the section then \emph{augments} the merged content -- that is, merged-in keys may be modified, and new keys may be introduced. These constructs can be nested (i.e., \ttinclude{}d documents can contain their own \ttinclude{}s) and intermixed (i.e., \ttuse\ can refer to sections defined by previously \ttinclude{}d content).

Further details are best left to the technical documentation -- here we rather provide a real-life illustration. Consider the case of CASA tasks. The \cultcargo\ package provides cab definitions for many CASA tasks, allowing them to be invoked individually from any step of the recipe. A number of these tasks (particularly, those that operate on visibility data) share a common collection of parameters which are used to specify the input measurement set, and data selection to be performed on that measurement set. Rather than repetitively listing these parameters within each cab's schema (which would be laborious and error-prone), we can define this collection of parameters once in a ``library'' file called {\tt casa.yml}: 

\begin{minted}[fontsize=\small]{yaml}
lib:
  params:
    casa:
      mssel_inputs:  
        # standard arguments for CASA tasks that take 
        # a vis argument, and a standard set of data 
        # selection options
        ms:
          info: Name of input visibility file
          dtype: MS
          writable: true
          required: true
          nom_de_guerre: vis
        field:
          info: Select field using field id(s) or 
            field name(s)
          dtype: str
        spw:
          info: Select spectral window/channels
          dtype: str
      #  ...
\end{minted}

We can then reuse this library content within any cab definition:

\begin{minted}[fontsize=\small]{yaml}
_include: casa.yml

cabs:
  casa.applycal:
    inputs:
      # pull in standard set of selection parameters
      _use: lib.params.casa.mssel_inputs
      # and now define some task-specific inputs:
      docallib:
        info: Use callib or traditional cal apply 
          parameters
        dtype: bool
      # ...
\end{minted}

The \ttinclude{} key may refer to a full path. Relative paths (and bare filenames) are resolved by looking in (a) the current directory, (b) the directory of the document invoking the include, (c) directories specified by the {\tt STIMELA\_INCLUDE} environment variable, and (d) some predefined locations such as {\tt ~/lib/stimela}, {\tt /usr/lib/stimela}, etc. A special form of the path allows one to refer to a Python package, e.g.

\begin{minted}[fontsize=\small]{yaml}
    _include: (cultcargo)wsclean.yml
\end{minted}    

\noindent will tell \stimela\ to look for {\tt wsclean.yml} within the installation directory of {\tt cultcargo}. 

While at most one \ttinclude\ key is allowed per section\footnote{A section is a YAML mapping, after all: the keys of a mapping must be unique.}, it is possible to include multiple files by providing a sequence (list) instead of a single path. The \ttinclude{} can also be structured if multiple files from the same location need to be pulled in:

\begin{minted}[fontsize=\small]{yaml}
    _include: 
      (cultcargo):
        - wsclean.yml
        - cubical.yml
\end{minted}

The applications of these constructs are plentiful and useful: libraries of recipes, libraries of step templates, reusing and augmenting step definitions within a recipe, etc. Programmers intuitively recoil at repetition, for both aesthetic and practical reasons -- it tends to promote hard-to-spot errors. The \ttuse{} and \ttinclude{} constructs can serve to avoid repetition and promote modularity within YAML documents. They are the key enabling technology that allows \stimela\ to work with distributable collections of cabs such as
\cultcargo.

\subsection{Composability}
\label{sec:composability}

The crucial concept enabled by the above constructs is that of \emph{composability}. Before it can run a workflow, \stimela\ needs to know many things, from the top-level recipe steps, to cab definitions, to the local (or cluster) environment configuration. Although it is perfectly possible to write a recipe file that provides all this information within a single big YAML document, this is hardly the recommended approach, for obvious reasons. Instead, one can leverage the use/include/merge/augment features described above to compose a workflow description from \emph{multiple} YAML files, which are ultimately all merged into the global configuration namespace. For example, a workflow such as that published by \citet{parrot} is composed of multiple documents:

\begin{itemize}
  \item The top-level recipe, describing the logic of the workflow (inputs, outputs and steps), provided by the recipe developer;
  \item Cab definitions for standard tools, provided by the \cultcargo\ package;
  \item Cab definitions for custom tools and scripts, provided by the recipe developer;
  \item Sub-recipes employed within the main recipe,  provided by the recipe developer;
  \item Local compute environment configuration (e.g. \kubernetes\ cluster setup), based on information provided by the system administrator;
  \item Optional additional configuration tweaks (e.g. to optimize performance), provided by the end user.
\end{itemize}

Composability promotes modularity and re-use of cab definitions and sub-recipes, and allows for a clear separation of responsibilities, particularly between the pure logic of the workflow, and the nuts-and-bolts specifics of the local compute environment. \ref{app:kubeconfig} provides an illustrative example of this.

\section{Cabs and cargo}

A {\bf cab} defines an atomic task available to \stimela, with  a defined set of inputs and outputs specified by a \emph{schema}.
The \emph{cargo} (content) of a cab can come in a variety of \emph{flavours}: 
an executable command, a function defined within a Python module, an individual CASA task, or even a snippet of inlined Python code.

\subsection{Cab definitions}

A cab definition is simply a section of YAML, Fig.~\ref{fig:cab} being an example. The illustrated cab will run the {\tt shadems} command. If a containerized backend is employed, it will use a container image called  {\tt stimela2/shadems}. 

The {\tt inputs} section defines the input parameter schema (see \sect{sec:schemas}). An optional {\tt defaults} section specifies default settings for parameters (alternatively, these can be embedded in the schema itself). The {\tt policies} section tells \stimela\ how to convert parameter values into command-line arguments (see \sect{sec:policies}). 

\subsection{Cab flavours}

The default cab flavour runs the specified executable command (natively, or within a container, depending on the configured backend), passing 
it arguments according to the specified policies. There are a few additional flavours available.

The {\tt python} flavour will invoke an arbitrary Python function, in which case the {\tt command} key specifies the function to be called as
as {\tt [\emph{package}.]\emph{module.function}}. The function's signature (i.e. arguments) must be correctly described by the cab schema. The function's return value can be interpreted as a cab output. Note that the function is invoked via an external Python interpreter (possibly in a separate virtual environment or a container).

The {\tt casa-task} flavour will invoke a CASA task. The {\tt command} key then specifies the task name. The task's signature (i.e. arguments) must be correctly described by the cab schema. 

The {\tt python-code} flavour can execute arbitrary Python code directly embedded into the cab definition. The {\tt command} key then specifies the code itself (YAML's use of the ``{\tt |}'' character to designate an indented multi-line string is particularly handy here). The cab's inputs are mapped to local variables before the code snippet is executed, and the outputs
are collected from local variables after the code snippet completes. This is useful for small tasks and ``glue code''. For example, the cab below will subtract the mean value from a FITS image, and return the resulting image and mean value as outputs:

\begin{minted}[fontsize=\small]{yaml}
cabs:
  subtract-mean:
    inputs:
      image:
        dtype: File
        required: true
    outputs:
      output-image:
        dtype: File
        required: true
      mean-value:
        dtype: float
    flavour: python-code
    command: |
      from astropy.io import fits
      ff = fits.open(image)
      mean_value = ff[0].data.mean()
      ff[0].data -= mean_value
      ff.writeto(output_image, overwrite=True)
\end{minted}

As an aside, the above example illustrates that YAML keys (and therefore cab names, parameter names, step labels, etc.) can contain dash characters (moreover, dashes within step names can be put to serve a useful purpose -- see \sect{sec:subst}.) Since Python variable names cannot contain dashes, \stimela\ will implicitly map them to underscores when communicating with a {\tt python-code} or {\tt casa-task} cab.
  
\subsection{Console output \& wranglers}

When running a cab, \stimela\ will intercept its console output and send it to its own logger (and console). There are cases when this output contains information that is required later in the workflow. For such cases, \stimela\ provides a mechanism called \emph{output wranglers}, which can parse and manipulate a cab's textual output in various ways, and capture its content into formal output parameters.

This is best illustrated by an example. Imagine that a workflow needs to know the fraction of visibilities flagged within a given measurement set. A short {\tt python-code} flavour cab can be readily defined, containing code to open the measurement set, compute this fraction, and return it in a cab output, as described above. An alternative is to invoke the CASA {\tt flagdata} task, ask it for a summary, and parse the value out of its output. (A more general version of this use case is exemplified by any external software package that prints some important item of informaton midway through its run -- how can we automatically capture and parse this information?) This can be done via the following cab definition:

\begin{minted}[fontsize=\small]{yaml}
cabs:
  flagsummary:
    command: flagdata
    flavour: casa-task
    inputs:
      ms: 
        dtype: MS
        required: true
        nom_de_guerre: vis
      mode:
        implicit: 'summary'
    outputs:
      percentage:
        dtype: float
    management:
      wranglers:
        'Total.* Counts: .* 
          \((?P<percentage>[\d.]+)%\)':
            - PARSE_OUTPUT:percentage:float
            - HIGHLIGHT:bold green
\end{minted}

The {\tt management.wranglers} section contains a sequence of regular expressions (employing standard Python {\tt re} module syntax), which are matched against each line of the output. A matching line then causes a number of actions to be invoked. In this case, the first action captures the flagged percentage (using a named regex group) and turns it into a {\tt float} value that is returned as the {\tt percentage} output of the cab. The second action highlights the matching line in bold green text, which is simply a cosmetic convenience for the user. 

Other useful wrangler actions can mark a cab as having succeeded or failed based on encountering some particular text output. By default, and as per normal Unix practice, \stimela\ determines success or failure via the return code of the underlying process. Certain software (notably, CASA) does {\bf not} return error codes on failure -- the wrangler feature allows \stimela\ to detect these error conditions anyway.

As another aside, note the {\tt implicit} definition for the {\tt mode} parameter of this cab. An implicit input is something that is not supplied by the user, but needs to be passed to the underlying cargo anyway. In this case, we pass {\tt mode='summary'} to the {\tt flagdata} task to force it into summary mode. (Likewise, implcit outputs can be used to describe the file-type outputs of a cab over the naming of which the user has no control.) 

\subsection{Cult-cargo and custom cargo}
\label{sec:cult}

The \cultcargo\ package is an (optional) companion to \stimela. This provides cab definitions for a number of standard radio astronomy packages, as well as for individual CASA tasks. \cultcargo\ is shipped through PyPI -- to use it, one only needs to {\tt pip install} it alongside \stimela. The cabs within can then be directly included into a recipe, as shown in \sect{sec:use-include}.

\cultcargo\ does not install the actual packages per se -- only their cab definitions. It does, however,
have an associated container image registry on {\tt https://quay.io}, where a collection of versioned images is maintained. References to these images are provided within the cab definitions. The upshot of this is that the end user need only specify that they want to use a containerized backend such as Apptainer/Singularity (\sect{sec:backends}) to use these images. Everything else is taken care of automatically by \stimela, making for a true installation-free workflow. On the other hand, developers or power users who want to run local custom installations of some (or all) packages are free to do so, by switching to the native backend. Note that it is also possible to specify backends on a per-cab (or per-step) basis, so a developer needing to experiment on some small part of the pipeline can temporarily switch to a native install for a particular step or cab, while continuing to use standard images everywhere else. This is done by augmenting the cab definition after including it:

\begin{minted}[fontsize=\small]{yaml}
  _include: (cultcargo)wsclean.yml
  cabs:
    wsclean:
      backend:
        select: native
      command: /path/to/my/wsclean
\end{minted}

The \cultcargo\ registry provides multiple image versions for each package. The default image is usually the latest stable release. However, if one wanted to experiment with a different version of the package for which \cultcargo\ provides an image, it trivially achieved by augmenting the definition with a specific version label:

\begin{minted}[fontsize=\small]{yaml}
  _include: (cultcargo)wsclean.yml
  cabs:
    wsclean:
      image:
        version: 2.10.1-kern7
\end{minted}
    
Developers can also provide customized cab definitions, for packages that have not (perhaps yet) made it into \cultcargo. For example, the PARROT recipes \citep{parrot} use some custom cabs from the {\tt omstimelation}\footnote{\url{https://github.com/o-smirnov/omstimelation}} collection. At time of writing, this is not a PyPI package, but rather just a git repository. To use it, once needs to clone the repository to a location in which \stimela\ knows to search for includes (\sect{sec:use-include}), and then invoke is as:

\begin{minted}[fontsize=\small]{yaml}
  _include: omstimelation/oms-cabs-cc.yml
\end{minted}

For custom cabs relying on container images, it is left up to the developers to set up an image registry and make their images available (and to specify their location in the cab definition) in order for their cabs to work with a containerized backend. The \cultcargo\ approach provides a distribution model that can be readily followed. Note that images do not need to be specified for {\tt python}, {\tt python-code} or {\tt casa-task} flavour cabs, since by default, these will use standard Python and/or CASA images provided by \cultcargo.

\section{Schemas}
\label{sec:schemas}

Schema definitions are at the core of \stimela's parameter definition and validation system. A schema defines the input and output parameters (the signature, in other 
words) of a cab or a recipe, and (optionally) a set of \emph{policies} that specify how the parameters are passed to the underlying cargo.

Some examples of schemas were already presented above, see e.g. \sect{sec:use-include}. Fundamentally, each cab or recipe definition contains an {\tt inputs} and/or an {\tt outputs} section, with named subsections defining individual parameters. These may be further nested if one wants to impose some kind of organizational 
hierarchy onto the parameters. For example, this schema

\begin{minted}[fontsize=\small]{yaml}
  inputs:
    data:
      ms:
        dtype: MS
        required: true
        info: Measurement set to process
      column: 
        dtype: str
        default: "DATA"
        info: column name
  outputs:
    image:
      dtype: File
      required: true
      info: output image
\end{minted}
  
...would specify two inputs, {\tt data.ms} and {\tt data.column}, and one \emph{named file} output, {\tt image}. (A named file output refers to the common situation 
where the user needs to specify the name of an output file or directory. Thus the file\emph{name} is, in some sense, an input -- however, the file itself is an 
output product, and is treated as such by \stimela. An alternative scenario is an {\em implicit} file output, where the user has no control over the filename, as 
it is determined by the underlying tool itself. This can be specified in the schema via an {\tt implicit} 
keyword supplying the filename.)

The {\tt dtype} entry specifies the type of the parameter, using the type annotation language 
provided by the {\tt typing} package of the Python Standard Library. Besides the built-in primitive types 
({\tt bool}, {\tt str}, {\tt int}, {\tt float} and the like), \stimela\ defines a few additional primitives ({\tt File}, {\tt MS}, {\tt Directory}). These can be 
combined with the annotation language to define rich types such as {\tt List[File]}, or even something like

\begin{minted}[fontsize=\small]{yaml}
  dtype: Union[str, Tuple[str, float]]
\end{minted}

\noindent which is useful for e.g. the imaging weight parameter, specifying something that can be either a single string (``uniform''), or a string and a number (``robust 0'').

Other schema options allow one to specify a parameter that can only assume a fixed set of values ({\tt choices}), to set a default value, to mark a parameter as {\tt required}, 
and to remap the name of the parameter when passing it to the underlying cargo ({\tt nom\_de\_guerre}).

\subsection{Shorthand schemas}

For those in a rush, \stimela\ also supports a shorthand schema syntax reminiscent of Python function signature annotations. The schema above can be expressed in shorthand as:

\begin{minted}[fontsize=\small]{yaml}
  inputs:
    data: MS * "Measurement set to process"
    column: str = "DATA" "column name"
  outputs:
    image: File * "output image"
\end{minted}

A shorthand schema contains the type, an optional default, an optional ``{\tt *}'' character indicating the parameter is required, and an optional trailing documentation string. To specify any other options, one must resort to a full schema definition as per the previous section. Shorthand and full schemas may be intermixed.

\subsection{Parameter validation}

The point of having well-defined schemas is to impose some robustness on the recipe definitions, and to catch errors as early as possible, rather than letting workflows fail midway through due to, say, missing parameters. \stimela\ accomplishes this in two stages. A {\em prevalidation} process is performed up front, before the recipe is executed. This checks that all required parameters
for all steps are defined, that the supplied types (of known parameters) are correct, that file-type inputs to the recipe are present (unless marked as {\tt must\_exist: false} in the schema), etc. The 
scope of pre-validation is naturally restricted -- after all, any non-trivial pipeline will have steps that depend on the outputs of preceding steps, which cannot be known upfront.
Further {\em runtime validation} is therefore performed before and after executing each step. This is a more thorough check ensuring that all the inputs to the step (which, by 
this point, are fully defined) match the schema. A similar check is done on the outputs of the step.

\subsection{Parameter policies}
\label{sec:policies}

Where the schema describes a (sub)recipe, or a cab with a Python function (or a CASA task) as its cargo, the information in the schema is necessary and sufficient for \stimela\ to invoke the cargo and pass arguments to it. If the cargo wraps a command-line tool, there is typically a one-to-one correspondence (modulo nom-de-guerre's) between the entries in the schema and the command-line parameters of the tool. However, different tools employ different command-line syntax conventions (a single dash prefixing options? A double dash? Are some arguments positional rather than optional? Does the tool rather use {\tt \emph{key=value}} for its arguments?) The optional {\tt policies} section of the schema is used to describe these argument-passing conventions. A default overall set of policies can be defined at cab level, and then redefined and tweaked at the level of individual parameters as needed. The variety of settings provided by the {\tt policies} section is sufficiently rich to represent most command-line conventions found in the wild. We leave details of this to the technical documentation.

\subsection{Dynamic schemas}

Some tools, \quartical\ \citep{africanus2}, and its predecessor \cubical\ \citep{cubical} being notable examples, have mutable command-line schemas, in the sense that the values of some command-line arguments
can affect the structure of following command-line arguments. For these two tools in particular, there is a ``Jones chain'' argument that specifies a list of Jones matrices to be included in the solution (e.g. G, K), in response to which the command-line parser starts recognizing extra arguments for specific settings related to these Jones terms (e.g. {\tt --g-type}, {\tt k-type}, etc.) \stimela\ supports this by providing a {\em dynamic schema} mechanism. A cab definition can include a callable Python function which takes in initial argument settings, and returns an updated schema for the cab. 

\subsection{Using schemas as a CLI builder}

For any given command-line tool, most of the information in the cab schema (i.e. argument names and types, help strings) directly mirrors that already provided to the tool's command-line parser. When wrapping a third-party package in a cab, this leads to an unavoidable duplication of effort (with all the attendant potential for inconsistencies) -- after all, the package developer has already implemented their own command-line interface (CLI) parser, and this CLI needs to be described to \stimela. Note, however, that the schema itself provides all the information that would be needed to construct a CLI in the first place. For newly-developed packages, this provides a substantial labour-saving opportunity. \stimela\ includes a utility function that can convert a schema into a CLI using the {\tt click}\footnote{\url{https://click.palletsprojects.com/}} package. For a notional example, consider this {\tt hello\_schema.yml} file defining a simple schema with two inputs:

\begin{minted}[fontsize=\small]{yaml}
inputs:
  name: 
    dtype: str
    info: Your name
    required: true
    policies:
      positional: true
      
  count:
    dtype: int
    default: 1
    info: Number of greetings
\end{minted}

This file can be instantly converted into a CLI as follows:

\begin{minted}[fontsize=\small]{yaml}
#!/usr/bin/env python
import click
from scabha.schema_utils import clickify_parameters

@click.command()
@clickify_parameters("hello_schema.yml")
def hello(count, name):
    """Simple program that greets NAME for a 
        total of COUNT times."""
    for x in range(count):
        print(f"Hello {name}!")

if __name__ == '__main__':
    hello()
\end{minted}

The resulting tool now has a fully-functional CLI:

\begin{minted}[fontsize=\small]{bash}
$ ./hello.py --help
Usage: hello.py [OPTIONS] NAME

  Simple program that greets NAME for a total 
  of COUNT times.

Options:
  --count INTEGER  Number of greetings
  --help           Show this message and exit.
\end{minted}

To integrate the tool into \stimela, all we need is a cab definition, which can directly include the schema file:

\begin{minted}[fontsize=\small]{yaml}
cabs:
  hello:
    _include: hello_schema.yml
    command: hello.py
\end{minted}

This mechanism ensures that all inputs and outputs need only be defined by the developer once, in a single place -- and provides both a CLI and \stimela\ integration with no additional effort, while ensuring that these 
are mutually consistent by construction. The \quartical\ and \pfbclean\ packages discussed earlier in this series \citep{africanus2,africanus3}, for example, make extensive use of this.

\section{Substitutions, evaluations, namespaces}
\label{sec:subst}

As briefly seen in \sect{sec:mainconcepts}, parameters of steps can refer to other parameters via constructs such as {\tt "=recipe.ms"} and {\tt "\{recipe.ms\}"}. These are examples of \stimela's more general substitution and evaluation mechanism, which is invoked when evaluating step parameters (as well as in a few other contexts, such as backend settings and logging options, see below.)

The mechanism has two basic rules: (a) a string that starts with an ``{\tt =}'' character is parsed using the formula syntax, and (b) all string values are subject to \{\}-substitution. Thus, the above two examples achieve identical results in very different ways.

In both cases, the underlying element (the variable in a formula, or the value being substituted in), {\tt recipe.ms}, is an instance of a \emph{namespace lookup}. 

\subsection{Namespace lookups}

The following namespaces are available for evaluation and substitution:

\begin{description}
    \item[\tt recipe] contains the parameters (inputs and outputs) and variables (see below) of the current recipe.
    \item[\tt current] contains the parameters of the current (i.e. ``this'') step in the recipe. A useful example would be:
    \begin{minted}[fontsize=\small]{yaml}
steps:
  convert-to-fits:
    image: =recipe.image
    output-fits: =STRIPEXT(current.image) + 
                 ".fits"
    \end{minted}
    which sets the output filename by stripping the extension from the input filename (using one of the predefined functions of the formula language, see below), and adding a different extension.
    \item[\tt previous] contains the parameters of the directly preceding step.
    \item[\tt root] contains the parameters of the top-level (root) recipe. 
    \item[\tt steps.foo] contains the parameters of (a necessarily preceding) step called ``foo''.
    \item[\tt info] contains information about the current step. For example {\tt info.label} is the step label, and {\tt info.suffix} is the \emph{suffix} of the label, if defined (that is, if the label contains dash characters, e.g. ``{\tt image-1}'', then the part following the last dash is the suffix, in this case ``{\tt 1}'').
    \item[\tt config] provides access to \stimela's entire config namespace (see \sect{sec:global-namespace}). For example, {\tt config.run.env.HOME} provides the value of the {\tt HOME} environment variable.
\end{description}

When doing namespace lookups, \stimela\ recognizes the wildcard character ``{\tt *}'' and interprets it in the sense of \emph{last matching element}. Thus, the following:

\begin{minted}[fontsize=\small]{yaml}
    image: =steps.make-image-*.output-image
\end{minted}

would refer to the {\tt output-image} parameter of the closest {\em preceding} step (preceding, because at any given step, only the preceding steps are present in the {\tt steps} namespace) whose label matches the {\tt make-image-*} pattern.

\subsection{The formula language}

Any parameter value starting with ``{\tt =}'' is parsed as a formula (unless it starts with ``{\tt ==}'', which evaluates to single equals sign, and treats the rest of the string as a literal). The formula language is patterned on basic Python syntax, and recognizes the following elements:

\begin{itemize}
    \item atomic elements, i.e. namespace lookups, as well as numeric and string constants.
    \item standard Python arithmetic and logical operators.
    \item the {\tt UNSET} keyword. If the formula evaluates to {\tt UNSET}, the corresponding parameter is deleted from the set of parameters passed to the step.
    \item a number of built-in functions, such as {\tt STRIPEXT()} above. Function names are always uppercase.
\end{itemize}

Please refer to the technical documentation for a complete list of available functions. Useful examples include:

\begin{itemize}
    \item Conditionals, such as
    
    {\tt IF(\emph{condition}, \emph{if\_true}, \emph{if\_false})}.
    
    \item Pathname manipulation functions, e.g. 
    
    {\tt STRIPEXT(\emph{path})}, {\tt DIRNAME(\emph{path})}

    \item Filesystem access, e.g.
    
    {\tt EXISTS(\emph{path})}, {\tt GLOB(\emph{pattern})}
    
    \item The {\tt NOSUBST(\emph{arg})} function, which evaluates is argument as a string, but then disables \{\}-substitutions on the result. With the exception of this function, \{\}-substitution is performed on all string elements (i.e. namespace lookups and string constants) within the formula.
\end{itemize}

\subsection{Curly brace substitutions}

The \{\}-substitution syntax is equivalent to Python format strings, since it uses that very mechanism under the hood. The braces must contain a valid namespace lookup, with an optional {\tt :\emph{format}} suffix to control the precise formatting of the string. 

Although both achieve the same means, there is a subtle difference between {\tt \{recipe.ms\}} and {\tt =recipe.ms}. The former does a namespace lookup and then formats the value as a string; the latter does a namespace lookup and returns the value directly. For string-type values, the result is the same, whereas for numeric-type values, the resulting data type will be different. (However, since \stimela\ knows to convert strings into numbers where a numeric parameter is defined in the schema, the actual outcome may end up being effectively the same anyway.)

\section{Recipe structure}

A recipe is defined in a YAML document that \stimela\ merges into its config namespace (\sect{sec:global-namespace}) on startup. As such, it can augment the content of any of the standard config sections, i.e. {\tt opts} to modify runtime options, {\tt cabs} to provide new cab definitions or to modify existing cabs, etc. Any top-level section in this document that doesn't match a config section is implicitly treated as a recipe body (and subsequently moved to {\tt lib.recipes}\footnote{Alternatively, recipes may be defined under {\tt lib.recipes} directly, but this adds two extra levels of indentation to the document.}.)

The recipe body consists of the following elements:

\begin{itemize}
    \item Optional {\tt name} and {\tt info} fields giving a recipe name and description;
    \item {\tt inputs} and {\tt outputs} sections containing schemas for the recipe's input and output parameters;
    \item An optional {\tt defaults} section specifying default parameter values (these can also be embedded directly in the schemas);
    \item An optional {\tt aliases} section (\sect{sec:aliases});
    \item Optional {\tt assign} and {\tt assign\_based\_on} sections specifying variable assignments (\sect{sec:assignments});
    \item An optional {\tt for\_loop} section which specifies that the recipe is a loop or a scatter-gather construct (\sect{sec:forloops});
    \item An optional {\tt backend} section to tweak recipe-specific backend settings (\sect{sec:backends});
    \item A {\tt steps} section specifying the sequence of steps.
\end{itemize}

\subsection{Step definitions}

Each step of the recipe is defined by a subsection under {\tt steps}. The key of each subsection is the step label, while the body of the subsection contains:

\begin{itemize}
    \item An optional {\tt info} field documenting the step.
    \item The cargo: this is either a {\tt cab} field specifying a cab, or a {\tt recipe} field specifying a sub-recipe. Normally, these are specified by name, but it is also possible to make this field into a subsection that directly embeds a full cab definition or sub-recipe definition.
    \item A {\tt params} section specifying the step parameters. 
    \item Optional {\tt assign} and {\tt assign\_based\_on} sections (\sect{sec:assignments}).
    \item An optional list of {\tt tags} associated with the step (\sect{sec:skips}).
    \item Optional {\tt skip} and {\tt skip\_if\_outputs} fields (\sect{sec:skips}).
    \item An optional {\tt backend} section to tweak step-specific backend settings (\sect{sec:backends});
\end{itemize}

The steps are executed in sequence. Before each step is run, \stimela\ performs evaluation and substitution on the parameters, matching them against the schema (the inputs) of the cargo. Any missing or invalid parameters cause execution to stop with an error. Likewise, after a step is run, its outputs are matched against the schema.

\subsection{Variable assignments}
\label{sec:assignments}

In addition to parameters, a recipe may define arbitrary variables (loosely speaking, the distinction is similar to that between the arguments and local variables of a function.) These become available for namespace lookup within the {\tt recipe} namespace. Variables are primarily useful for setting up parameter substitutions and enforcing things such as filename conventions. They are defined via the {\tt assign} and {\tt assign\_based\_on} sections. The concept is best illustrated by a real-life example. Imagine that we have a number of measurement sets, and we want to associate an ``observation label'' with each one, so that (a) the input MS, and a number of associated settings, can be selected via a single {\tt obs=\emph{label}} parameter to the recipe, (b) all outputs of the workflow go into a subdirectory called {\tt ./obs-\emph{label}}, and (c) the outputs of each imaging step are separated into subdirectories within. This can be achieved by structuring the recipe as follows:

\begin{minted}[fontsize=\small]{yaml}
my-recipe:
  inputs:
    obs:
      dtype: str
      info: label of observation
      required: true
    ms:
      dtype: MS
  outputs:
    dir.out:
      dtype: Directory
      default: obs-{recipe.obs}
  assign:
    image-prefix: {recipe.dir.out}/im{info.suffix}/
                  im{info.suffix}
  assign_based_on:
    obs:
      L1:
        ms: ms1-l.ms
        band: L
      L2:
        ms: ms2-l.ms
        band: L
      U1:
        ms: ms1-u.ms
        band: UHF
    band:
      L:
        image-size: 6000
        image-scale: 1arcsec
      UHF:
        image-size: 6000
        image-scale: 2arcsec
  steps:
    # ...
    image-1:
      cab: wsclean
      params:
        ms: =recipe.ms 
        name: =recipe.image-prefix
        scale: =recipe.image-scale
        size: =recipe.image-size
    ...
\end{minted}

When this is run with {\tt obs=L1}, \stimela\ proceeds as follows:

\begin{itemize}
    \item The {\tt dir.out} parameter receives a default value of {\tt obs-L1}.
    \item The {\tt image-prefix} variable is assigned to via the {\tt assign} section (the \{\}-substitutions within are unresolved at this point, but this is not considered an error until the variable is used elsewhere, by which point the substitutions become valid.)
    \item Finding an {\tt obs} section under {\tt assign\_based\_on}, \stimela\ looks up the key {\tt L1} within that section, and performs assignments based on its content. The {\tt ms} parameter is assigned to (unless a value has explicitly been specified by the user on the command line), and a new variable called {\tt band} is defined and assigned the value {\tt "L"}.
    \item The latter assignment has further repercussions: finding a {\tt band} section under {\tt assign\_based\_on}, \stimela\ looks up {\tt "L"}, and assigns {\tt image-size} and {\tt image-scale} accordingly.
\end{itemize}

The assignment sections are then re-evaluated before every step. When \stimela\ gets to the {\tt image-1} step, the {\tt info.suffix} lookup results in {\tt "1"}, so the {\tt image-prefix} variable is set to {\tt "obs-L1/im1/im1"}. \wsclean\ is then invoked with that as its output name prefix, and scale/size parameters appropriate to the selected band.

Notice that the {\tt assign\_based\_on} section here contains fairly observation-specific information. It is probably bad practice to hardwire it into the recipe body itself, as that mixes specific configuration with generic recipe logic. A cleaner solution is to split out specific configuration into a separate YAML document, and either include it (\sect{sec:use-include}) into the recipe via

\begin{minted}[fontsize=\small]{yaml}
  assign_based_on:
    _include: obs-config.yml
\end{minted}

\noindent or specify it as a separate document on the command line. \stimela\ itself does not prefer and will not enforce any particular convention here, it merely provides the tools to cleanly separate generic logic from specific configurations in a number of different ways.

\subsection{Aliases}
\label{sec:aliases}

Often, a recipe parameter is directly passed to a step as is. In the example above, this is achieved via a {\tt =recipe.ms} evaluation. \emph{Aliases} provide a way to formalize this relationship. Instead of specifying an evaluation (or substitution) at the step level, one could remove the definition of the recipe's {\tt ms} input, and add an {\tt aliases} section instead:

\begin{minted}[fontsize=\small]{yaml}
    aliases:
        ms:
        - image-1.ms
\end{minted}

This tells \stimela\ that the recipe has an input (or output) named {\tt ms} with the exact same schema as the {\tt ms} input of step {\tt image-1}, and that the value of {\tt ms} should be passed to that step directly. The {\tt ms: =recipe.ms} setting at step level is then obviated. The difference between an alias and an evaluation or substitution is that an alias is a more formal, hard link -- it implicitly declares a matching schema at recipe level, and the input parameter can be checked up front before the recipe is run (whereas {\tt =recipe.ms} is only evaluated before the step itself is run, so no errors can be caught until then.) An alias can link a recipe parameter to multiple steps:

\begin{minted}[fontsize=\small]{yaml}
    aliases:
        ms:
        - image-1.ms
        - calibrate-1.input-ms.path
        - image-2.ms
\end{minted}

The above can also be specified more economically via wildcards:

\begin{minted}[fontsize=\small]{yaml}
    aliases:
        ms:
        - image-*.ms
        - calibrate-*.input_ms.path
\end{minted}

\noindent or by referring to all instances of a particular cab:

\begin{minted}[fontsize=\small]{yaml}
    aliases:
        ms:
        - (wsclean).ms
        - (quartical).input_ms.path
\end{minted}

In certain situations, aliases are generated automatically. For example, if the recipe contains an {\tt image} step referring to a cab with a mandatory {\tt ms} parameter, but a value for {\tt ms} is not provided within the step body, \stimela\ will automatically generate a recipe-level alias named {\tt image.ms}, which then becomes a required recipe parameter. More subtly, any \emph{optional} step parameters that are not explicitly set in the step are also auto-aliased to recipe-level parameters (using the same naming scheme). \stimela\ considers such recipe parameters ``obscure'' or ``hidden'' (depending on whether the parameter has a default), since the recipe can be operated without the user needing to know anything about them. However, these parameters are still accessible from the command line (and can also be set when the recipe is invoked as a sub-recipe), should the user need to tweak them. The {\tt stimela doc} command does not list obscure or hidden parameters by default, but provides an option for it (see {\tt stimela doc {-}{-}help}).

\subsection{For-loops}
\label{sec:forloops}

A recipe can be turned into a loop by providing a {\tt for\_loop} section:

\begin{minted}[fontsize=\small]{yaml}
    for_loop:
        var: counter
        over: [0,1,2]   # or equivalently, =RANGE(3)
\end{minted}

This declares a recipe-level {\tt counter} variable, and iterates the recipe three times, for the three different values of {\tt counter}. Instead of providing a list of values directly, the {\tt over} field can also refer to another recipe variable or input: 

\begin{minted}[fontsize=\small]{yaml}
    my-recipe:
        inputs:
            ms-list:
                dtype: List[MS]
        for_loop:
            var: ms
            over: ms-list
            scatter: 4
\end{minted}

The {\tt scatter: 4} setting tells \stimela\ to run up to four iterations of the loop in parallel (use {\tt -1} to run all iterations in parallel). If a distributed backend such as \kubernetes\ or Slurm is configured (\sect{sec:backends}), this will distribute jobs across the cluster (and if not, all four jobs will run on the local machine -- it is up to the user to ensure that this has enough resources to support the parallelism.)

Note that \stimela's composability features (\sect{sec:composability}) make it trivial to turn a non-looping recipe into a looping one. Let's say we have a {\tt recipe.yml} document defining a (non-looping) recipe called {\tt my-recipe}, with an {\tt ms} input. By placing the YAML snippet above into a separate document called {\tt loop-recipe.yml}, and running

{\small
\begin{verbatim}
$ stimela run recipe.yml loop-recipe.yml 
    "ms-list==GLOB('*.ms')"
\end{verbatim}
}

\noindent we augment our recipe into a for-loop over all matching measurement sets.

\subsection{Skips and tags}
\label{sec:skips}

The optional {\tt skip} field of a step can be used to tell \stimela\ to skip the step. A simple {\tt skip: true} setting is what's called a force-skip -- normally, it is the equivalent of commenting the step out (note, however, that the {\tt -s} option to {\tt stimela run} can be used to override the skip and execute the step anyway.) More interestingly, one can declare a \emph{conditional} skip based on a formula evaluation. For example,

\begin{minted}[fontsize=\small]{yaml}
    skip: =EXISTS(current.output-file)
\end{minted}

\noindent will tell \stimela\ to skip the step if its {\tt output-file} parameter refers to an existing file. This is such a common use case in workflows with expensive steps that \stimela\ supports an explicit {\tt skip\_if\_outputs} field:

\begin{description}
    \item[\tt skip\_if\_outputs: exist] will cause a step to be skipped if all its file- and directory-type outputs exist.
    \item[\tt skip\_if\_outputs: fresh] will cause a step to be skipped if all its file- and directory-type outputs are ``fresh'', in the sense that their modification times are newer than or equal to that of the most recent file- or directory-type input. Those familiar with Unix-style Makefiles will readily recognize this logic.
\end{description}

The optional {\tt tags} field serves a related purpose, and can be used to group related steps of a long workflow into subsets that can be selected or deselected for execution \emph{en masse}\footnote{The tag selection logic is inspired by and modelled on Ansible playbooks (\url{https://www.ansible.com}).}. In a nutshell, invoking 

{\small
\begin{verbatim}
$ stimela run recipe.yml -t foo
\end{verbatim}
}

\noindent will run only the steps that have a {\tt foo} (and/or {\tt always} tag -- the latter having this special meaning), and skip all others. A second special tag is {\tt never}. Steps tagged with {\tt never} are normally skipped, unless they have another tag that has been explicitly specified with {\tt -t}.

\section{Backends}

\label{sec:backends}

\stimela's \emph{backend} components are responsible for actually scheduling and executing the jobs (cabs). At time of writing, the following backends are supported:

\paragraph*{Native} This backend runs the command (or Python interpreter, or CASA) natively on the host system. Optionally, a separate virtual environment may be specified for any given cab or step. The native backend provides the most flexibility in a development or experimental environment, but the onus is entirely on the user to make sure that all the underlying packages are installed and available on the host. Consequently, this offers the least reproducibility -- running the recipe on another system requires that exactly the same versions of all software packages be installed.

\paragraph*{Singularity/Apptainer} This backend runs the command (or Python interpreter, or CASA) using the Apptainer\footnote{\url{https://apptainer.org}. Apptainer was formerly branded as Singularity, and \stimela's current documentation, as well as the configuration syntax, still uses the old name in some contexts. Older, Singularity-branded versions of the engine (version 2.6+) are currently still supported. The old name will likely be deprecated in future versions of \stimela.} containerization engine. The image from which the container is instantiated can be downloaded automatically from a container registry (such as {\tt quay.io}, where the standard \cultcargo\ images are maintained, see \sect{sec:cult}) and built on the spot, or may be provided by the user as a local SIF (Singularity image format) file. At present, Apptainer is the container engine of choice for HPC environments\footnote{Docker and Podman provide similar functionality, and backends for these engines can probably be added to \stimela\ in the future, given sufficient demand.}.

The Apptainer backend allows for true zero-install, fully reproducible workflows. The host system only requires an installation of \stimela\ and Singularity/Apptainer, while all required images are downloaded on-demand (but can also be pre-downloaded by the user). The host system must, of course, have sufficient resources (primarily, RAM and disk) to accommodate all the steps of the workflow.

\paragraph*{Kkubernetes} This backend runs the command (or Python interpreter, or CASA) on a \kubernetes\ cluster in a {\em  pod}, using a container associated with a registry. From the user's point of view, the \kubernetes\ backend is also zero-install -- there is, however, considerable onus on the system administrator to configure and provide access to a \kubernetes\ cluster. \kubernetes\ has emerged as the technology of choice for scalable workflows, particularly cloud-based ones. The results shown earlier in this paper series use the \stimela\ \kubernetes\ backend to run workflows on the AWS platform. This backend will be described in some detail below. 

\paragraph*{Slurm} This backend (or strictly speaking, backend wrapper) can schedule jobs remotely via the \slurm\ scheduler, using the native or Apptainer backends to actually run the jobs on the compute nodes.

\paragraph*{Specifying backends} Backends are specified via a {\tt backend} section in four possible locations: (a) the top-level {\tt opts} config section, (b) the recipe definition, (c) the cab definition, and (d) a particular step definition. At each step, \stimela\ will compose the settings from the four locations in the order given here, and apply the resulting settings. In a routine end-user workflow, only the top-level backend (a) is specified -- this can be as simple as adding

\begin{minted}[fontsize=\small]{yaml}
  opts:
    backend:
      select: singularity
\end{minted}

\noindent to the recipe file so as to run the entire workflow using Apptainer/Singularity. Options (b) through (d) are meant for development workflows, as well as tuning resource allocation under \slurm\ or \kubernetes. In principle, they even allow a workflow to mix-and-match backends. For example, one could quickly tweak the recipe to switch from a standard \wsclean\ \cultcargo\ image to a locally-built binary. This allows for rapid experimentation, but certainly does not promote reproducibility.

\subsection{The Slurm backend wrapper}

The \slurm\ wrapper must be combined with the native or Apptainer/Singularity backend. It can be enabled as follows:

\begin{minted}[fontsize=\small]{yaml}
  opts:
    backend:
      select: singularity
      slurm:
        enable: true
\end{minted}

In a \slurm\ deployment, \stimela\ itself is executed on the head (login) node, since it is a relatively lightweight process requiring very little resources. It then wraps job invocations in \slurm's {\tt srun} command, which causes them to be executed on compute nodes, and captures the jobs' console output via its logging mechanism, while waiting for jobs to complete. This approach naturally dovetails with the scatter feature of loop recipes (\sect{sec:forloops}), and makes it straightforward to deploy parallel workflows across the cluster.

The {\tt srun} command has a wide array of options to control resource allocation (e.g. CPUs, RAM) and node placement. Instead of actively managing these, \stimela\ takes a hands-off approach, providing a transparent mechanism for tuning {\tt srun} on a per-cab and per-step basis. An optional {\tt srun\_opts} subsection under {\tt backends.slurm} is mapped to {\tt srun} options -- that is, every {\tt \emph{key}: \emph{value}} entry in that subsection is converted into {\tt -{}-\emph{key} \emph{value}} arguments on the {\tt srun} command line. Since backend options are composed hierarchically (see above), this allows for a fine degree of tuning, e.g.:

\begin{minted}[fontsize=\small]{yaml}
  opts:
    backend:
      select: singularity
      slurm:
        enable: true

  cabs:
    wsclean:
      backend:
        slurm:
          srun_opts:
            cpus-per-task: 32
            mem: 128G

  recipe:
    steps:
      foo:
        backend:
          slurm:
            srun_opts:
              cpus-per-task: 1
              mem: 16G
      bar:
        backend:
          slurm:
            srun_opts:
              cpus-per-task: 4
              mem: 32G
\end{minted}

\stimela's composability features (\sect{sec:composability}) allow for this tuning information to be placed into a separate YAML document that can be combined with the recipe itself at runtime.

\subsection{The Kubernetes backend}

\kubernetes, often denoted as \knetes, is a highly extensible, open-source container orchestration platform designed to facilitate the deployment, scaling, and management of distributed, containerized applications. Emerging from Google, based on its experience of running containers at scale, \kubernetes\ has rapidly become the de facto standard in container orchestration, achieving widespread adoption in the cloud-native computing community. \kubernetes\ is a cross-platform standard, and provides a unified and abstracted interface to a ``cluster'' resource that is largely independent of the underlying implementation, the latter taking many forms and scales, e.g.:

\begin{itemize}
\item Elastic \kubernetes\ Service (EKS) provided by AWS;
\item Google \kubernetes\ Engine (GKE) provided by Google Cloud;
\item Azure \kubernetes\ Service;
\item On-premises cluster solutions such as {\sc Rancher}\footnote{\url{https://rancher.com}} and the {\sc MicroK8s}\footnote{\url{https://microk8s.io}} engine;
\item A private ``virtual cluster'' on the local machine, provided by the \kubernetes\ In Docker ({\sc kind})\footnote{\url{https://kind.sigs.k8s.io}} package.
\end{itemize}

The \stimela\ \kubernetes\ backend is implemented in terms of the standard \knetes\ Python API. While the recipe itself is parsed and managed by the local \stimela\ installation, each actual step of the recipe is executed as a {\em pod} on a (potentially, remote) \knetes\ cluster -- while console output and logfiles are captured locally. This means that both the logical structure of the recipe, as well as the end-user look-and-feel, remain the same for both local and cloud-based\footnote{At time of writing, we have only tested this with the AWS EKS cloud implementation. However, the backend -- by design -- knows nothing about AWS per se, with only the standard \knetes\ API is employed throughout. We can thus reasonably expect that other cloud architectures can be supported with minimal to no effort. The onus is entirely on the cluster administrator to configure a K8s cluster on the appopriate platform.} workflows (modulo, of course, the data itself being made available on the relevant platform.)

The \kubernetes\ backend tends to require a lot more configuration (see \ref{app:kubeconfig} for an example), with input from relevant system administrators. Fundamentally, this is because the \knetes\ paradigm tends to deal in highly abstracted resources, managed via {\em operators}, which in turn manage the lifecycle of these resources. For example, instead of a single (local or networked) POSIX-like filesystem (as would generally be the case in a local or \slurm\ environment), \knetes\ deals in {\em volumes} and {\em persistent volume claims}, which are mapped onto underlying cluster resources. \stimela\ does not offer any features to simplify \knetes\ cluster administration -- the cluster must be preconfigured, and this requires specialist administrator knowledge. However, once this is in place, \stimela's  composability features (\sect{sec:composability}) allow for a clean separation of roles and responsibilities:

\begin{itemize}
\item The cluster administrator sets up and maintains the \knetes\ environment, as well as access rights to the cluster. Ultimately, the users are instructed on how to configure a \knetes\ {\em context} in which their jobs will be executed, and how to obtain cluster access credentials.

\item The context is specified in the {\tt opts.backend.kube} section (in fact, the context may be omitted, if the user has selected a default context in their \knetes\ configuration file).

\item Additional information from the administrator is used to configure recipe-specific resources, such as volumes and node allocations, in the {\tt opts.backends.kube} section.

\item All this configuration can be defined in a separate YAML document, which exists independently of the recipe.

\end{itemize}

In principle, a clean separation can be achieved, such that the difference between running a \slurm\ and a \knetes\ workflow can be as little as 

{\small
\begin{verbatim} 
  $ stimela run recipe.yml slurm-config.yml 
\end{verbatim}
}
\noindent versus
{\small
\begin{verbatim}
  $ stimela run recipe.yml kube-config.yml 
\end{verbatim}
}

\subsection{Resource allocation}

When running a recipe via a local backend, \stimela\ simply executes the steps one-by-one (or in parallel, if a loop scatter construct is employed), and defers to the operating system to enforce resource limits (via disk quotas, cgroups, etc.) However, an optional {\tt backends.rlimits} subsection can be configured to set per-process limits in the \stimela\ session. This supports all the standard {\tt RLIMIT\_} symbols defined by the PSL {\tt resource} module\footnote{\url{https://docs.python.org/3/library/resource.html}}.

When it comes to resource \emph{requests} (as opposed to limits), e.g.  the number of CPU cores to use, different packages have different conventions for specifying such. It is generally left up to the recipe developer to propagate these options to the packages appropriately. A good practice, for example, is to define a recipe-level input:

\begin{minted}[fontsize=\small]{yaml}
  inputs:
    ncpu:
      dtype: int
      info: max number of CPUs to use
      default: =config.run.ncpu-physical
\end{minted}

\noindent ...and pass this to cabs consistently. Note that the {\tt config.run} namespace provides information on the number of physical and logical cores on the local system.

In a \slurm\ or \kubernetes\ environment, resource management becomes somewhat more elaborate. \stimela\ does not attempt to control this directly. Instead, the philosophy is to expose the relevant options via backend settings (mirroring the options of the underlying cluster interface as much as possible, rather than trying to invent an extra abstraction layer), and let the user tweak these on a per-cab and per-step basis using hierarchical composition. We saw an example of this with the {\tt srun\_opts} section of the \slurm\ backend wrapper above. In the \kubernetes\ environment, the standard \knetes\ API provides the concept of CPU and memory \emph{requests} and \emph{limits}, which can be controlled via the 
{\tt backend.kube.job\_pod.cpu} and {\tt backend.kube.job\_pod.memory} subsections. To give another example, platform-specific implementations of \knetes\ provide fine-grained control over where a pod is scheduled (for example, on what kind of AWS EC2 instance) via a custom {\tt nodeSelector} section in the pod spec. \stimela\ simply exposes this section under {\tt backend.kube.job\_pod}, for the user to tune (see \ref{app:kubeconfig}).

\subsection{Custom resource management}

\newcommand{\dasknetes}{{\sc Dask-kubernetes}}

One exception to \stimela's hands-off resource policy is the management of \dask\ jobs on \knetes, controlled via the \dasknetes\footnote{\url{https://kubernetes.dask.org/}} operator. The \knetes\ API supports the concept of \emph{custom resources} (CRs), defined by \emph{custom resource definitions} (CRDs). \dasknetes\ uses this API to define a \textsc{DaskJob} CRD, hierarchically composed of \textsc{DaskCluster} and \textsc{DaskWorkerGroup} CRDs, which include definitions for \emph{job runner}, \emph{scheduler} and \emph{worker} pods. Upon creation of a \textsc{DaskJob} CR, the operator creates all the dependent CRs in a hierarchical fashion, bringing up the pods. The operator then monitors the execution of the job, retrying the three pod types on failure. Upon successful completion of the job, or after a limited number of failures, the operator destroys the pods. Explicit deletion of the \textsc{DaskJob} results in the operator deleting all related resources.

To support \dask-aware application such as \quartical, \stimela\ provides a {\tt backends.kube.dask\_cluster} section, where the components of the \textsc{DaskJob} are defined. As with all other backend settings, this can be tweaked on a per-cab or even per-step basis. The content of the section is passed on to the \dasknetes\ API. On completion (or failure) of the \textsc{DaskJob}, \stimela\ deletes it, ensuring that all related CRs are released.

A second exception is temporary disk storage, required by some applications (consider, e.g., \wsclean\ and its {\tt -temp-dir} option). With local and \slurm\ backends, this is simply a temporary directory within the filesystem, so no particular management is required. In the \knetes\ environment, a volume and a persistent volume claim must be configured, and must be cleaned up properly to avoid hogging a (potentially costly) resource. \stimela\ provides control over this via the {\tt backends.kube.volumes} section; temporary volumes can be configured to persist for the duration of the recipe ({\tt lifecycle: session}) or even just the step ({\tt lifecycle: step}).

\section{Diagnosing complex workflows}

As anyone familiar with developing pipelines (or even simple processing scripts) will know, two of the most common questions that arise in the process are (a) what went wrong? and (b) what took so long? \stimela\ provides a couple of mechanisms to help provide [at least first-order] answers.

\subsection{Logfile management}
\label{sec:logs}

For a complex workflow with many steps, a full output log, while usually extensive and chaotic, can yet be an invaluable
debugging aid. \stimela\ tries to aid this further. When running a recipe, it will intercept each cab's console (stdout and stderr) output, and send a copy to its log (as well as to its own console). It then provides a number of features to organize log files in a sensible and human-friendly way; these are invoked by assigning to the {\tt opts.log} section of the config namespace. This can be done directly within a recipe's YAML code. Here's an example of generically useful logger settings (employing the substitutions described in \sect{sec:subst}):

\begin{minted}[fontsize=\small]{yaml}
opts:
  log:
    dir: logs/log-{config.run.datetime}
    name: log-{info.fqname}
    nest: 2
    symlink: log
\end{minted}

This snippet will configure \stimela\ to do the following:

\begin{itemize}
\item The logfiles for each run are stored under a unique log subdirectory named {\tt ./logs/log-\emph{YYYYMMDD-HHMMSS}/}, where the timestamp refers to the date/time the (outer) recipe was started.
\item Each step of the recipe is logged into its own separate logfile, based on its fully-qualified name (i.e., {\tt log-\emph{recipe\_name}.\emph{step\_label}.txt}).
\item The output of nested sub-recipes, if any, is logged as part of the enclosing step (this is implied by nesting level 2). Increasing the nesting level will cause nested sub-recipes to generate their own uniquely-named logfiles.
\item The symlink {\tt logs/log} is updated to point to the most recent log subdirectory.
\end{itemize}

\subsection{Profiling}

\stimela\ includes some basic profiling functionality. When using a local (native or Apptainer/Singularity) backend, it 
collects and reports the following set of performance metrics:

\begin{itemize}
    \item Elapsed time
    \item CPU use percentage
    \item RAM use
    \item System load
    \item Read/write operations (Gb per second)
    \item Read/write operations total.
\end{itemize}

These metrics are broken down by step (hierarchically, if sub-recipes are employed), as well as by peak and average, and reported to the console at the end of a run, as well as saved to a YAML file for future analysis. This allows for quick identification of both particularly slow steps, as well as resource-hogging ones.

For more detailed profiling of individual steps, it is a simple matter to modify the cab definition so as to invoke an external profiler, for example:

\begin{minted}[fontsize=\small]{yaml}
cabs:
  wsclean:
    command: valgrind --tool=callgrind wsclean
\end{minted}

With the \slurm\ backend wrapper, \stimela\ can (currently) only measure elapsed time.

With the \knetes\ backend, \stimela\ uses the \knetes\ metrics API to collect metrics from running pods. However, since the detailed metrics provided by this API are highly platform-dependent, \stimela\ is currently restricted to only the basic set of standard metrics:

\begin{itemize}
    \item CPU core usage
    \item RAM usage
    \item Number of running pods
    \item Elapsed time (for which no API is required, of course).
\end{itemize}

The ultimate profiling metric (when running on the cloud) would be dollars per run (or dollars per MeerKAT image). In principle, Amazon's EKS engine, at least, seems to provide the necessary custom APIs to make collecting this information feasible. We have deliberately avoided incorporating any vendor-specific APIs into the \kubernetes\ backend, but a possible avenue for future development is adding a ``plugin'' capability to support such custom metrics.

\section{Conclusions, discussion and future work}

We believe \stimela\ has largely achieved its aim of occupying the middle ground between ease of use (linear scripting with a rich syntax), scalable workflows (provided by the \slurm\ and \kubernetes\ backends), and practical reproducibility (provided by containerization). It has also enabled us to run workflows in the cloud (as demonstrated by the use of AWS for some of the benchmarks in this paper series.) It is already used as the underlying platform for some of our in-house projects \citep[see e.g.][]{solarkat}. \stimela\ has enabled extremely non-standard data reductions \citep{parrot}, with the latter work providing an important example of publishing a science paper along with reproducible recipes. We would welcome and support wider adoption of \stimela\ in the community. \stimela\ is fully open source, and has a stable public release available via PyPI.

This paper series presents a new software ecosystem that is already reasonably feature-complete: it is now almost feasible to run an entire MeerKAT data reduction, from raw visibilities to final images, within a \stimela\ workflow, using exclusively \daskms-based packages (implying that we can dispense with a CTDS-backed measurement set and use a more efficient, parallel-I/O-enabled storage backend, throughout). At the same time, \stimela\ remains fully compatible with legacy tools, the PARROT recipes \citep{parrot} being a case in point -- these readily fall back on CASA and \wsclean\ when needed. This development raises a number of interesting issues meriting further discussion.

\subsection{Whither cloud?}

Our new software ecosystem removes two of the technical bottlenecks to adopting cloud solutions in radio astronomy. Firstly, the \daskms\ data access layer allows us to replace the traditional CTDS storage backend with modern \zarr- or \apache-based backends that can use S3 object stores, the most economical cloud storage solution which also scales linearly via requests on multiple nodes. This also opens the door to exploring interesting price/performance optimizations, since cloud providers such as AWS offer hierarchical object store solutions with progressively pricier/cheaper faster/slower storage.

Two related points should be made here:

\begin{itemize}
    \item One does not need to be all-in on \daskms\ to start exploiting the cloud. We also use \stimela\ to run cloud workflows that intermix legacy packages such as CASA and \wsclean\ with our new-generation tools. To support access via legacy packages, the data then necessarily has to reside in CTDS-backed measurement sets on block storage. This will continue to remain an option going forward, albeit a less cost-effective and performant one.
    \item The flexibility provided by the \quartical\ and \pfbclean\ packages can further reduce storage costs (of selfcal-style workflows), by (a) obviating the need for separate model and corrected data columns (in a throw-back to the traditional AIPS approach, a \quartical\ to \pfbclean\ selfcal loop need only store the raw data, plus per-antenna gain solutions), and (b) providing support for baseline-dependent averaging.
\end{itemize}

Secondly, \stimela, and its \kubernetes\ backend in particular, shows a way to resolve the ``thick-thin'' problem of traditional workflows. Pods are scheduled with predefined CPU and RAM requirements. The \knetes\ autoscaler is then able to bring the required number (and type) of virtual machine instances up and down on demand, with ``thin'' steps allocated to small and cheap instances, and large expensive instances only spun up when ``thick'' steps can make effective use of them.

We don't claim to have solved all the problems of cloud computing for radio astronomy here, and neither are we (yet) in a position to claim that the cloud is the more (or less) cost-effective solution for our needs -- although the example of Rubin Observatory looms large. In particular, we have completely ignored the issue of data ingress and egress. What we can claim is that (a) our new software ecosystem does enable more economical workflows (in a cloud context) than the legacy software stack, and that (b) it does open the door to quite detailed price/performance evaluations and optimizations in the future. (As an aside, we further note that cost analysis on the cloud is a fairly simple and transparent process -- it is much more intricate for on-premises compute, where power, labour costs, etc. need to be factored in.)

Finally, in a research software context, the cloud does offer an absolutely unique opportunity for algorithm evaluation. At any given point in time, any on-premises cluster (or HPC centre) is locked into a particular set of hardware configurations and architectures. If one wanted to evaluate an algorithm's performance on a different architecture (how does my solution scale to 100 thin nodes? Can I make efficient use of 16 GPUs at the same time?), these resources may simply not be available locally, or even at national HPC centres. Cloud providers, on the other hand, offer a rich variety of architectures for short-term rental. This comes with a price tag, of course -- but having an option with a price tag is better than having no option at all. We therefore believe that cloud computing will have a large role to play in our community going forward, regardless of its adoption by major radio telescopes.

As an anecdotal data point to illustrate this, the total cost for this paper series (and associated software packages), in terms of AWS cloud charges, came to approximately \$25,000 over two years. This included a lot of software development and testing, many abortive, misguided and simply ill-conceived experiments, and the final, often massively scaled, benchmark runs presented in the previous papers. This is comparable to the price of a single high-end compute node, without power and administration costs. While a compute node has a useful lifetime of well over two years, even a high-end node would not have accommodated some of the scaling experiments presented in this series. As another data point, we recently conducted a four-day data reduction workshop (Africalim 2024\footnote{\url{https://github.com/africalim/resources}}) with $\sim50$ participants, introducing them to the Africanus software stack. Virtually all the data tutorials were run on an AWS EKS cluster, with participants spinning up multiple instances of workflows at the same time, without experiencing any resource contention. The total cloud charges for the workshop came in at under \$1,000.

\subsection{Wither reproducibility?}

\stimela's combination of fully containerized workflows, as well as the maintenance of versioned images on \cultcargo's {\tt quay.io} registry, means that any given recipe should be able to be run reproducibly across a range of architectures (natively, on \slurm, or even \kubernetes), modulo availability of CPU and RAM resources. However, this still needs to be proven in practice. In terms of the reproducibility tenets defined by \citet{daliuge-repr}, \stimela\ is designed to provide \emph{scientifically replicatable} workflows. In terms of \emph{computational replication}, there are subtle caveats. Numerical algorithms are subject to round-off errors, may behave differently on different architectures, and even give subtly different results when parallel computation results in the order of operations not being guaranteed. \emph{Total replicability} may always remain hostage to the robustness of algorithm implementations.

There is also a sociological aspect to the reproducibility issue. \stimela\ quite deliberately makes development workflows easy to implement: test images and locally-built binaries may be swapped in by changing one line in a YAML file. This makes it tempting to violate reproducibility, in the name of quicker experimental turnaround -- the onus remains on the user to provide the self-discipline when it comes to publishing results that have been obtained with reproducible workflows. 

In order to encourage such self-discipline, it will be worthwhile to introduce the concept of \emph{certifiable} versus \emph{non-certifiable} workflows. A certifiable \stimela\ workflow is one that operates fully in containerized mode, while \emph{exclusively} using versioned images from public registries. Any other workflow is then non-certifiable, and can't be guaranteed to be reproducible. The current version of \stimela\ takes some steps in this direction, by automatically generating a dependencies file listing all images and software versions (including its own) employed in the workflow. With a little more development effort, and adopting some of the ideas of \citet{daliuge-repr}, this can become the basis of a formal certification mechanism. Container images already provide signature hashes. These can be combined with a hash of the full YAML configuration, and perhaps even a hash of the input data, to produce a \emph{certified workflow signature}. Anyone attempting to reproduce the workflow can then, as a minimum, be guaranteed that they're starting from 
bit-perfect copies of the input data and software packages. 

\subsection{A public recipe competition}

As an important signpost in this direction, we intend to shortly release a set of end-to-end \stimela\ recipes that take some (publicly available) challenging MeerKAT datasets all the way from raw visibilities to final images. The raw data will be hosted on AWS via its Open Data program\footnote{\url{https://aws.amazon.com/opendata}} (and is, in any case, also available from the SARAO archive.) In the first instance, this will allow the community to test our reproducibility claims.

More importantly, this will offer an interesting opportunity to \emph{provably} compare calibration and imaging algorithms. Radio interferometry literature is full of any number of attempted ``imaging competitions'', as well as many papers illustrating the superior imaging performance of the authors' novel algorithm, via a comparison with best-effort images made with a competing package. When presented in paper form, it is inevitable that the impact of such comparisons is limited and often contentious. Ideally, what we would like to see instead are public releases of modified recipes where, for example, the imaging step has been swapped out for the authors' novel package. If such recipes are made publicly available, and can be run reproducibly, then algorithm comparisons become far more meaningful and quantitative. We therefore intend to promote our end-to-end recipes as the start of an open \emph{recipe} (as opposed to \emph{image}\/) competition.

\section*{Acknowledgements} 

Funding: OMS's and JSK's research is supported by the South African Research Chairs Initiative of the Department of Science, Technology and Innovation and National Research Foundation (grant No. 81737).

SM's research is supported by a {\it SARAO-Funded University Research Groups} grant of the South African Radio Observatory and the National Research Foundation (grant No. 97792).

The MeerKAT telescope is operated by the South African Radio Astronomy Observatory, which is a facility of the National Research Foundation, an agency of the Department of Science, Technology and Innovation. 

We would like to thank Ross Davies, Ilze van Tonder, Christopher Voges, Peter Fosseus, Nawaal Adams, Dietrich Keller and their colleagues at Silicon Overdrive, as well as Agnat Max Makgoale (Amazon Web Services) for invaluable technical and logistical assistance over the course of this work.

\paragraph*{Data Availability} The MeerKAT data used for the benchmarks in this paper series
is publicly available via the SARAO archive\footnote{\url{https://archive.sarao.ac.za}}, see \citet{africanus2}, \cite{africanus3} for details.





\bibliographystyle{elsarticle-harv} 

\bibliography{stimela} 




\appendix

\section{Basic recipe walk-through: Simulating a MeerKAT observation}
\label{app:telsim}

This example shows a recipe that will create a simulated measurement set containing visibility data products, based 
on a source catalogue. This simulation follows the following steps:

\begin{enumerate}
    \item Configuration of the telescope and observing parameters, such as:
    \begin{itemize}
        \item Observing frequency, channelization and the number of channels.
        \item Observation start time, integration time and duration.
        \item Phase centre.
    \end{itemize}
    \item Definition of the sky to be observed, and simulation of visibility data. 
    \item Imaging of the simulated data.
\end{enumerate}

Now, let's write the corresponding \stimela\ recipe. This simulation will use the \telsim\ and \skysim\ tools from the \simms\ package\footnote{\url{https://github.com/wits-cfa/simms}}, as well as the \wsclean\ cab from the standard \cultcargo\ suite. This assumes that \simms\ and \cultcargo\ have been installed (e.g. via pip).

\begin{minted}[fontsize=\small]{yaml}
_include:
  - (simms.parser_config)simms-cabs.yaml
  - (cultcargo)wsclean.yaml
\end{minted}

The \telsim\ tool creates an observation template (or empty MS) given an array telescope configuration and observing parameters, and \skysim\ simulates a sky model into an MS. Both these tools are loaded in the {\tt \_include} section from the {\tt simms-cabs.yaml} configuration file that comes with the {\tt simms} package. We will also use the \wsclean\ package, for which we pull in a cab definition from the standard \cultcargo\ suite.

Next, we set the runtime options for the recipe. Here, we set the recipe to use the host's native software environment (i.e. native backend).
This assumes that all the software (including the \wsclean\ binary) is installed locally.
We also direct all log files into a directory {\tt logdir} which will nest the logs for each run in a subdirectory labelled with the run's timestamp (\sect{sec:logs}). Logs from the latest run can be accessed via a shortcut (symlink) named {\tt logs}:

\begin{minted}[fontsize=\small]{yaml}
opts:
  backend:
    select: native
  log:
    dir: logdir/logs-{config.run.datetime} 
    nest: 2
    symlink: logs
\end{minted}

Now that the recipe dependencies and runtime settings are done, let's write the recipe per se. We start with defining the recipe inputs (recall that {\tt dtype} is {\tt str} by default):

\begin{minted}[fontsize=\small]{yaml}
simulator-recipe:
  info: "Basic simulation example"
  inputs:
    prefix:
      default: example-simulation
    msname:
      implicit: "{current.prefix}.ms"
    telescope:
      default: meerkat
    direction:
      default: J2000,0deg,-30deg
    dtime:
      dtype: float
      default: 10
      info: "correlator dump time, in seconds"
    ntimes:
      dtype: int
      default:  180
      info: "number of correlator dumps"
    freq0:
      dtype: Union[float,str]
      default: 1.4GHz
      info: "reference frequency"
    dfreq:
      dtype: Union[float,str]
      default: 1MHz
      info: "channel width"
    nchan:
      dtype: int
      default: 5
      info: "number of channels"
    skymodel:
      dtype: File
      required: yes
    sefd:
      dtype: float
      default: 420
      info: "telescope SEFD"
    npix:
      dtype: int
      default: 4096
      info: "image size in pixels"
    pix-size:
      default: 1.2asec
      info: "pixel size"
\end{minted}

These inputs can be changed at runtime. For example, to run the simulation for the VLA telescope's C configuration instead of the default MeerKAT telescope, the command line would be

\begin{minted}{bash}
  $ stimela run telsim.yaml telescope=vla-c
\end{minted}

Now, we add the simulation outputs. These are (a) the simulated MS and (b) the imaging products. For the latter,
we use aliases (\sect{sec:aliases}) to directly propagate out the outputs of the imaging step:

\begin{minted}[fontsize=\small]{yaml}
  outputs:
    ms:
      dtype: MS
      implicit: =current.msname
    image:
      aliases: [image.restored.mfs]
    dirty-image:
      aliases: [image.dirty.mfs]
    resid-image:
      aliases: [image.residual.mfs]
    psf-image:
      aliases: [image.psf.mfs]
\end{minted}
  
Finally, we add the three steps described at the beginning of this section. \noindent The first step, {\tt makems}, uses \telsim\ to create an empty MS using the defined observation settings. The settings defined in the {\tt recipe.inputs} section can be accessed as recipe attributes, as will be shown in this step. The output of this step is the empty MS.
  
\begin{minted}[fontsize=\small]{yaml}
  steps:
    makems:
      cab: telsim
      params:
        ms: =recipe.msname
        telescope: =recipe.telescope
        direction: =recipe.direction
        dtime: =recipe.dtime
        ntimes: =recipe.ntimes
        startfreq: =recipe.freq0
        dfreq: =recipe.dfreq
        nchan: =recipe.nchan
\end{minted}

The second step, {\tt addsky}, uses \skysim\ to simulate the sky model into the MS. This step updates the {\tt DATA} column of the MS with the simulated visibilities, and does not create any new files.

\begin{minted}[fontsize=\small]{yaml}
    addsky:
      cab: skysim
      params:
        ms: =steps.makems.ms
        catalogue: =recipe.skymodel
        sefd: =recipe.sefd
        column: DATA
\end{minted}

The last step, {\tt image}, uses \wsclean\ to make an image of the simulated observation.

\begin{minted}[fontsize=\small]{yaml}
    image:
      cab: wsclean
      params:
        ms: =steps.makems.ms
        prefix: =recipe.prefix
        column: =steps.addsky.column
        size: =recipe.npix
        weight: briggs 0
        niter: 10000
        scale: =recipe.pix-size
        mgain: 0.85
        auto-mask: 5
        auto-threshold: 2  
\end{minted}

This last step yields a number of FITS files, some of which are propagated to the recipe's outputs via the aliases defined above.

The full recipe can now be invoked as

\begin{minted}{bash}
  $ stimela run telsim.yaml 
\end{minted}

As noted above, any non-default recipe inputs can be added to the command line.

\onecolumn

\section{Configuring a mixed workflow including Kubernetes}
\label{app:kubeconfig}

The YAML code below provides a practical example of advanced \stimela\ usage, illustrating several key points.

\begin{itemize}
    \item This code is not a standalone recipe, but is rather an augmentation of the base recipe 
    presented in Appendix B of \citet{africanus3}. It is invoked together with the base recipe, as e.g.:

    {\small
    \begin{verbatim}
    $ stimela run pfbimage.yaml kubeconfig.yaml image obs=esohi basedir=s3://rarg-test-binface/ESO137 \
          log-directory=/mnt/data/pfb-test/logs
    \end{verbatim}}

    \item The purpose of the augmentation is to test and benchmark the base recipe. See comments in YAML code for more details.

    \item The augmentation results in most of the recipe being executed on AWS, but the first step is configured to execute locally using the 
    Singularity/Apptainer backend.
    
    \item The augmentation also adds two additional steps at the end of the base recipe, also executed locally. Cab 
    definitions for these steps are provided below. 

    \item The code also contains a complete example of configuring the \kubernetes\ backend for an AWS EKS cluster.
\end{itemize}

Altogether, this illustrates how (rather complicated) runtime deployment logic and performance tweaks can be kept completely separate from the logical structure of the recipe.

\begin{minted}{yaml}
opts:
  backend:
    select: kube
    # set up global kubernetes config 
    kube:
      enable: true
      dir: /mnt/data/pfb-test
      namespace: rarg-test-compute
      volumes:
        rarg-test-compute-efs-pvc:
          mount: /mnt/data

      user:
        uid: 1000
        gid: 1000

      # Some predefined pod specs to be utilised in the recipe
      predefined_pod_specs:
        admin:
          nodeSelector:
            rarg/node-class: admin
        scheduler:
          nodeSelector:
            rarg/node-class: compute
            rarg/instance-type: m6i.large
            rarg/capacity-type: ON_DEMAND
        medium:
          nodeSelector:
            rarg/node-class: compute
            rarg/instance-type: c6in.8xlarge
            rarg/capacity-type: ON_DEMAND
        large:
          nodeSelector:
            rarg/node-class: compute
            rarg/instance-type: c6in.24xlarge
            rarg/capacity-type: ON_DEMAND

      # common environment variables for worker nodes
      env:
        NUMBA_CACHE_DIR: /mnt/data/pfb-test
        CONFIGURATT_CACHE_DIR: /mnt/data/pfb-test
        JAX_ENABLE_X64: 'True'
        JAX_PLATFORMS: 'cpu'
        LD_LIBRARY_PATH: '/usr/local/lib'

image:
  info:
    This file augments the stimela config in the main imaging recipe.
    It demonstrates the use of multiple backends, including the kubernetes
    backend which enables processing some of the steps on AWS instances. 
  steps:
    init:
      info: 
        Reads data from local storage and writes the averaged
        Stokes visibility products, potentially to an S3 bucket
        associated with an AWS account.
      backend:
        select: singularity
        singularity:
          rebuild: true
          bind_dirs:
            # These need to be mounted inside the singularity image
            # Substitute $USER for bester here
            /home/bester/numba_cache: rw
            /home/bester/.aws: rw
          env:
            NUMBA_CACHE_DIR: /home/bester/numba_cache

    grid:
      info:
        Sets up a Dask cluster that distributes gridding tasks per band. 
      backend:
        select: kube
        kube:
          enable: true
          # The main application runs on the job_pod
          job_pod:
            type: medium
            memory:
              limit: "54Gi"
            cpu:
              # request more than half the available cores
              # to utilize the full instance per pod
              request: 20
          # Spawns Dask cluster with 6 worker nodes 
          dask_cluster:
            enable: true
            num_workers: 6
            name: pfb-test-cluster
            threads_per_worker: 1
            worker_pod:
              type: medium
              memory:
                limit: "55Gi"
              cpu:
                request: 20
          # Environment variables can be adjusted per step
          env:
            NUMBA_NUM_THREADS: '32'
      params:
        fits-mfs: false
        fits-cubes: false
        nworkers: 6
        nthreads: 32


    sara:
      info:
        The main deconvolution application runs on a single large node.
      backend:
        select: kube
        kube:
          enable: true
          job_pod:
            type: large
            memory:
              limit: "170Gi"
            cpu:
              request: 50
          dask_cluster:
            enable: false
      params:
        # FITS files could be written to EFS but not S3.
        # They are not required for the test we perform below.
        fits-mfs: false
        fits-cubes: false
        nworkers: 1
        nthreads: 96

    pull_model:
      info:
        Transfer component model from S3 to local storage for testing 
      cab: s3tolocal
      # binary command is executed on the native backend
      backend:
        select: native
      params:
        source: s3://rarg-test-data/hi_combined_bda_I_main_model.mds
        dest: /home/bester/projects/ESO137/from_aws/hi_combined_bda_I_main_model.mds

    compare_models:
      info:
        Ensure that the model produced on AWS infrastructure is compatible
        with the model produced locally up to deconvolution tolerance.
      cab: compmodels
      # python script executed on the native backend in a
      # virtual environment that has pfb-imaging installed
      backend:
        select: native
        native:
          virtual_env: ~/.venv/pfb
      params:
        model_local: /scratch/bester/hi_combined_bda_I_main_model.mds
        model_remote: =previous.dest
        epsilon: =recipe.steps.sara.tol

cabs:
  s3tolocal:
    name: s3tolocal
    info: 
      Runs a binary command to transfer data from AWS S3 to local storage. 
    command: aws s3 cp --recursive
    policies:
      positional: true
    inputs:
      source:
        info: The location to fetch the model from
        dtype: URI
        required: true

    outputs:
      dest:
        info: The location to place the model
        required: true
        dtype: Directory
        mkdir: true

  compmodels:
    name: compmodels
    info: 
      Runs a python script to test whether two component model are
      compatible up to a tolerance specified by epsilon. 
    flavour: python-code
    command: |
      import numpy as np
      import xarray as xr
      from pfb.utils.misc import model_from_mds

      model1 = model_from_mds(model_local)
      model2 = model_from_mds(model_remote)
      model1 -= model2  # diff
      model1 += 1.0     # for relative tolerance
      try:
        assert np.allclose(model1, 1.0, atol=epsilon, rtol=epsilon)
      except:
        raise ValueError("Models don't match")

      print('Success!')

    inputs:
      model_local:
        info: Model obtained by executing locally
        required: true
        dtype: str
      model_remote:
        info: Model obtained by executing remotely
        required: true
        dtype: str
      epsilon:
        info: Precision with which to compare models
        default: 1e-4
        dtype: float
\end{minted}

\end{document}